\documentclass[usenatbib]{mn2e}
\usepackage{graphicx}
\usepackage{longtable}
\usepackage{color}

\font\hf = cmsl7 scaled \magstep 0
\title[The effects of UV photometry and binary interactions on photo-$z$ and galaxy morphology]
{The effects of UV photometry and binary interactions on photometric redshift and galaxy morphology}

\author[F. Zhang et al.]
{Fenghui~Zhang$^1$\thanks{E-mail: gssephd@public.km.yn.cn;
zhang\_fh@hotmail.com}, Zhanwen~Han$^1$, Lifang~Li$^1$, Hongguang~Shan$^1$ and Yu~Zhang$^{1,2}$\\
$^1$National Astronomical Observatories/Yunnan Observatory, Chinese
Academy of Sciences, Kunming, 650011, China\\
$^2$Graduate University of the Chinese Academy of Science, Beijing
100049, China}

\begin{document}

\date{\today}

\pagerange{\pageref{firstpage}--\pageref{lastpage}}

\pubyear{2010}

\maketitle

\label{firstpage}

\begin{abstract}
Using the \textit{Hyperz} code (Bolzonella et al. 2000) and a template spectral library which consists of 4 observed galaxy spectra from Coleman, Wu \& Weedman (CWW, 1980) and 8 spectral families built with evolutionary population synthesis models, we present photometric redshift estimates { (photo-$z$)} for a spectroscopic sample of 6,531 galaxies, for which spectroscopic redshifts are also available and are selected from the SDSS DR7 and GALEX DR4, and morphologies for a morphological sample of 1,502 bright galaxies, which are from the catalogue of Fukugita et al. { (2007)} and also matched with the SDSS DR7 and GALEX DR4.

{ We find that the inclusion of $F_{\rm _{UV}}$ or $N_{\rm _{UV}}$ or both photometry decreases the number of catastrophic identifications ($|z_{\rm phot}-z_{\rm spec}| > 1.0$).
If catastrophic identifications are removed, the inclusion of both $F_{\rm _{UV}}$ and $N_{\rm _{UV}}$ photometry mainly increases the number of non-catastrophic identifications in the low redshift, $g-r \la 0.8$ and fainter $r$-magnitude regions.
The inclusion of binary interactions mainly increases the number of non-catastrophic identifications and decreases the deviations in the $0.3 \le g-r \le 0.8$ region in the case of only using optical photometry.}

{ Based on the morphological galaxy sample, we find that the inclusion of UV photometry would decrease and increase the probability that early types are classified as Burst and E types, respectively, and increase that late types are classified as CWW-Sbc and CWW-Scd types.
If catastrophic identifications are excluded, the inclusion of UV data mainly raises the identifications of late types in all redshift, bluer $g-r$ and $r \ga 14$ regions. Moreover,
Binary interactions mainly affect the determinations of E and S0 types.}

{ By comparison we find that the {\sl reliability} and {\sl completeness} for early- and late-type selection by the \textit{Hyperz} code are less than those by the concentration index $C=2.6$, profile likelihood $P_{\rm exp}-P_{\rm deV}=0$ and colour $u-r=2.22$ criteria.
Moreover, we find that $N_{\rm _{UV}}-u = 1.94$ and $5.77-1.47(u-r) = F_{\rm _{UV}}-u$ discriminators can be used as morphology selection indicators. These two criteria have comparable {\sl reliability} and {\sl completeness} for selecting early- and late-type galaxies to $C=2.6$ criterion and higher {\sl completeness} for early-type selection than $u-r=2.22$ criterion.
}
\end{abstract}


\section{Introduction}
Redshift is one of the key ingredients { of} cosmology, it can be used to estimate distances and hence to place observed properties on a physical scale{ . M}any modern astrophysical measurements and researches (identifications of galaxy { clusters} and very high redshift objects, baryon acoustic oscillation measurements, galaxy evolution, large scale structure and gravitational lensing studies) { have benefited} from substantially larger catalogues of redshifts \citep{ger09}.

Galaxy redshift plays an important role in studies of galaxy formation and evolution. { Up to now, the spectroscopic redshifts of many relatively bright galaxies are obtained, while those of fainter galaxies are difficult to obtain.} Because that photometry is $\sim$ 2 orders of magnitude lesser time-consuming than spectroscopy for a given telescope size, the redshifts of fainter galaxies ($R \ga 25${\,mag}) or upcoming surveys will exclusively rely on the broad-band, medium-band or custom-designed narrow-band photometry \citep*[][and references therein]{bra08,nie09}.

{ The technique} for deriving redshifts from broad-band photometry { was} pioneered by \citet{bau62}. So far, a number of codes { have been developed} to compute photometric redshifts (photo-$z$s), such as {\sl BPZ} \citep{ben00}, {\sl Hyperz} { \citep*{bol00}}, {\sl GREGZ} \footnote{Because Greg Rudnick did not name his code, we follow the nomenclature of \citet{bra08}} \citep{rud01,rud03}, {\sl I{\hf MP}Z} \citep{bab04}, {\sl Le\,PHARE} \footnote{See http://www.oamp.fr/people/arnouts/LE\_PHARE.html.}(Arnouts \& Ilbert), {\sl ZEBRA} \citep{fed06}, {\sl kcorrect} \citep{bla07}, {\sl LRT} \citep{ass08}, {\sl EAZY} \citep{bra08}, {\sl ARBORZ} \citep{ger09}, Artificial Neural Networks (ANNs) and so on.
{ Three kinds of algorithm have been used in the photo-$z$ calculations} \citep{abd08,bra08,ger09,nie09,zha09}{ .}
\begin{itemize}
\item {T}he template fitting approach { has even been used. I}n this scheme photo-$z$ is obtained by comparing the observed spectra { with} a set of template spectra, which can be based { on either} population synthesis models or empirical studies.
\item {T}he empirical training set approach { has been used. T}he basic principle of this method is the derivation of a parameterization of redshift through the magnitudes of the galaxies in a training set, this parameterization is then applied to galaxies whose redshifts need to be estimated, yielding an estimation of the photometric redshift. The galaxies in the training set have known spectroscopic redshifts and similar properties to the galaxies for which one wants to estimate the redshifts. { ANNs is an} empirical training set approach.
\item { I}nstance-based learning method { also has been used. This method} also relies on the real data.
\end{itemize}

Among the traditional template fitting codes{, some of them} (such as {\sl Hyperz}, {\sl I{\hf MP}Z}, {\sl Le\,PHARE}, etc) used a straightforward $\chi^2$ minimization algorithm to obtain photo-$z$.
As spectral information provided by broadband photometry is at low resolution, { the photo-$z$s of galaxies have relatively large errors.} To avoid catastrophic errors, some codes improved the template spectra, for example, \citet{nie09} developed a set of high-resolution spectral templates based on { galaxy} physical information about the star-formation history{ , other} codes used a refinement of the $\chi^2$ minimization method{, for example,}
(i) { {\sl BPZ} code incorporated Bayesian statistics and allowed the use of extra information of galaxies (priors),
(ii) {\sl EASY}, {\sl ZEBRA} and {\sl kcorrect} codes used the hybrid method:}
{\sl EASY} code allowed the linear combination of templates as done in {\sl GREGZ} code and the use of priors;
{\sl ZEBRA} code used the combination of the priors { [}which were used to calculate a prior self-consistently from the photometric catalogue when it is run in Bayesian mode{ ]} with the training set method { [}which was used to improve the standard set of templates in specified redshift bins in its template optimization mode by the advantage of the zCOSMOS database \citep{lil07}{ ]}; {\sl kcorrect} code used the data of Sloan Digital Sky Survey \citep[SDSS,][]{yor00} project and incorporated the principle component analysis method to optimize the template.
{ Among these photo-$z$ codes, the} priors of luminosity function \citep{mob07} and surface luminosity \citep{xia09} have been used.

{ Among those non-traditional photometry fitting codes, some of them} included the structural properties of galaxies, for example, \citet{wra08} used surface brightness and the S\'ersic index (a measurement of the radial light profile) in addition to five-band photometry to obtain photo-$z$s{, and} \citet{sar99} used bulge-to-total flux ratio along with $I$-magnitude and $V-I$ colour to obtain photo-$z$s{.} \citet{neg09} { incorporated} the features of polycyclic aromatic hydrocarbons and silicate features into mid-infrared wavelength window to recover the photo-$z$s of starburst galaxies.
{ Moreover,} \citet[$\mu$-photoZ]{kur07} even only used one colour and the surface brightness from a single band to obtain photo-$z$s of galaxies.

Now, photo-$z$s have been extensively { used in} many studies, such as { they were used} in deep cosmological surveys to yield the galaxy luminosity function and the evolution of star formation { rate (SFR)}\citep[][and references therein]{mob07}.

\vskip 0.5cm
Morphology is another important parameter of galaxy. { First}, it is one of the first segregation of galaxy clusters, different morphological types exhibit distinctly different astrophysical properties, reflecting the different histories of the formation and evolution of galaxies \citep{shi01}. Studying it can help us to investigate the properties of classified galaxies [such as colours, the absence/presence of emission lines or the galaxy profiles \citep{lin08}]. Moreover, studying the correlation of morphology with other galaxy properties (such as luminosity function, \citealt{bla01}) can help us to understand the physical processes in galaxies (such as the influence of star formation on the process of galaxy formation) and the theories of galaxy formation and evolution \citep{bla01}.
So far, several approaches and techniques have been proposed to determine the morphological types{ .}
\begin{itemize}
\item { A}utomated classification approach, i.e., { using} the parameter(s) sensitive to morphology to obtain automatically the galaxy morphology{, has been used}. { This method spends less time, but faces on a serious problem that it is difficult to find such parameter(s).} In history, the following parameters have been used in the automated classification method: $u-r$ colour  \citep{str01}, profile probabilities (the relation between exponential $P_{\rm exp}$ and de Vaucouleurs' $P_{\rm deV}$ profile likelihoods), concentration index \citep[the ratio of the radii containing 90\% to 50\% of the \textit{Petrosian} $r$ galaxy light, $C\equiv r_{\rm p90}/r_{\rm p50}$,][]{shi01,str01} and so on.
\item { V}isual classification approach, i.e., { classifying} galaxies by visual inspection of experts { or} volunteers{, has been used}. This method serves as the most reliable method when adopting the Hubble classification \citep{san61} for galaxies with large apparent sizes \citep{shi01}. The greatest disadvantage of this method is more time-consuming, the size of the galaxy sample is small and the galaxies inspected are relatively bright.
    \citet{fuk07} have compiled a catalogue about morphological classification of $\sim2500$ bright galaxies in the SDSS Data Release three (DR3) by the visual inspection of three expert classifiers.
    Galaxy Zoo project has invited a large number of volunteers to separate early-type galaxies from spirals in { the} large data sets by using proxies for morphology \citep{lin08}.
\item { S}pectrum fitting approach, i.e., fitting the observed spectra to the template spectra of galaxies with known morphological types{, has been used}.
\item { T}he $k-$means method { also has been used} \citep{zha09}.
\end{itemize}

\vskip 0.5cm
We will use one of the widely adopted template-fitting codes, \textit{Hyperz}, to obtain the photo-$z$s of galaxies{ . M}eanwhile the rough morphological types will also be obtained because the \textit{Hyperz} code { allows to obtain} the best set of fit parameters.
As the first step of this work, we will use { Hyperz code to} obtain the photo-$z$s and morphological types { of galaxies} with known spectroscopic redshifts and morphological types{. This} can be used to explore the ability of the \textit{Hyperz} code of recovering redshifts and morphological types of galaxies, and judge whether or how the recovered parameters { are} affected by { the} input parameters {(including ultraviolet [UV] photometry, binary interactions and population synthesis models)}.

The reasons of discussing the influence of UV { photometry} and binary interactions on the determinations of photo-$z$s and morphological types { are as follows.}
\begin{enumerate}
\item { In} those previous works using template fitting method to obtain photo-$z$s of galaxies, the template spectra { usually} came from { real objects or population synthesis models of single stellar populations (SSPs), rather than} the population synthesis models including binary interactions.
    As for population synthesis models { of binary stellar populations (BSPs)}, { the} Yunnan group \citep{zha04b,zha05} have considered { various} binary interactions in evolutionary population synthesis (EPS) models, and drawn the conclusion that the inclusion of binary interactions can affect the overall shape of the spectral energy distribution (SED) of population, in particular, the SED of population in the UV passbands is bluer by 2-3\,mag at $t \sim 1$\,Gyr if  binary interactions are accounted for.
    Above conclusion further has been confirmed by \citet[][hereafter HPL07]{han07}, who included hot subdwarf B stars (sdBs) in population.
    The bluer SED in the UV passbands {  can be explained by that} binary interactions can create some important classes of objects, { such as sdBs} for { a} population older than $\sim$1\,Gyr (HPL07) and blue stragglers at 0.5$\sim$1.5\,Gyr.
\item { Observations find} that almost all elliptical galaxies have UV-upturn phenomenon, i.e., the flux increases with the wavelength decreasing from 2000 to 1200\,$\rm \AA$ (HPL07). { This} phenomenon can be explained by the EPS models { of BSPs (HPL07)}. Thus, if only optical photometry and/or the EPS models { of SSPs are used} to obtain photo-$z$s and morphologies of elliptical galaxies, whether these results are reasonable or not?

\item Another fact is that the photo-$z$s, which { are} derived by using the \textit{Hyperz} code under the condition of neglecting UV photometry, significantly deviate from the spectroscopic ones for some galaxies in the redshift range $z<0.2$ (also see Fig.~\ref{Fig:lowz-dia1}).
\end{enumerate}

{ In this work,} we use two samples of galaxies: spectroscopic and morphological. The galaxies in the spectroscopic sample are selected randomly from the SDSS DR7 and their spectroscopic redshifts are available. The galaxies in the morphological sample are from the catalogue of \citet{fuk07}, which presented the morphological types of 2,253 bright galaxies in the SDSS DR3 by using independent classification scheme (visual classification). All galaxies in both samples have matched with the Galaxy Evolution Explorer (GALEX) DR4 (see { \color{red} Section} 3 for details).
{ Moreover, we not only use the population synthesis models of SSPs but also those of BSPs to construct theoretical template library (see Section 2).}
\citet{nie09} have used the SDSS and GALEX photometric data to { study} the redshifts of galaxies, but they used the population { synthesis} models of { SSPs}.

The outline of the paper is as follows. In Sections 2 and 3 we describe the method  { used} to obtain photo-$z$ and the galaxy samples, respectively. In Sections 4 and 5 we obtain photo-$z$ estimates for the spectroscopic galaxy sample and morphologies for the morphological galaxy { sample}, and discuss the effects of UV photometry and binary interactions on them. Finally we present a summary and conclusions in Section 6.

\section{Method}
\begin{table*}
\centering \caption{Description about GISSEL98, BC03, HPL07 and Yunnan
{ spectral synthesis} models.}
\begin{tabular}{rlrl}
\hline
        &GISSEL98\&BC03        &HPL07  &Yunnan \\
{ population}     & { (SSP)}                 &{ (SSP\&BSP)} & { (SSP\&BSP)} \\
Ages                           &  221 &   88     &    90 \\
\hline
${\rm log}(t_i/{\rm yr})$      &  {5.100} &  { 8.000}   &   {5.000}\\
${\rm log}(t_f/{\rm yr})$      & {10.300} & 10.175   &  10.175 \\
$\Delta{\rm log}(t/{\rm yr})$  & {0.050}\ \ \ \ \ \ \ \ \ \  [{5.100} $\le
{\rm log}(t/{\rm yr}) \le$ {6.000}; $7.757 \le {\rm log}(t/{\rm yr}) \le
9.207$]
 &  0.025  &  {0.100} [{5.000} $\le {\rm log}(t/{\rm yr}) \le$ {6.500}] \\
                               & {0.020}\ \ \ \ \ \ \ \ \ \  [{6.000} $\le {\rm log}(t/{\rm yr}) \le$ {7.440}]
 &         &  {0.050} [${\rm log}(t/{\rm yr}) \ge${ 6.500}]\\
                               & 0.005$\sim$0.038 [{7.440} $\le {\rm log}(t/{\rm yr}) \le 7.757$; $ {\rm log}(t/{\rm yr}) \ge 9.207$] &          &         \\
\hline
\end{tabular}
\label{Tab:ages}
\end{table*}

\begin{figure}
\centering{
\includegraphics[bb=78 35 582 703,height=7.0cm,width=7.0cm,clip,angle=-90]{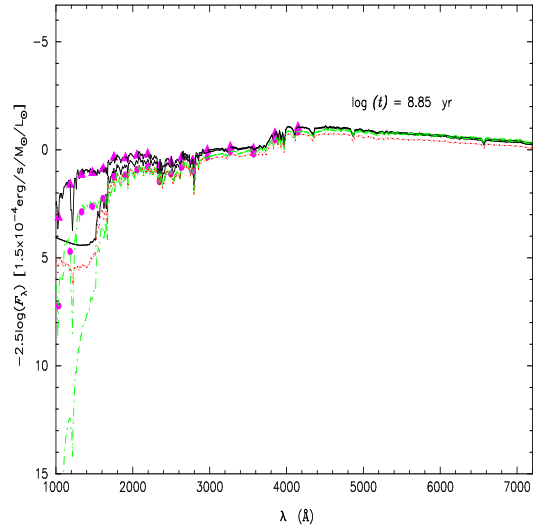}
}
\caption{Comparison of SED among BC03 { (red dotted line)}, Yunnan (black { solid} line + triangles for Yunnan-b; black { solid} line for Yunnan-s) and HPL07 models (green { dot-dashed} line + circles for HPL07-b; green { dot-dashed} line for HPL07-s) for a solar-metallicity population ({ normalized to}1\,${\rm M_\odot}$) at ${\rm log} (t) = 8.85$\,yr.} \label{Fig:isedcom-ath}
\end{figure}

The photo-$z$ and galaxy morphology are computed through the \textit{Hyperz} code of \citet{bol00}, which is the first publicly available photo-$z$ code and has been widely used in the literature for photo-$z$ { estimations of galaxies} \citep{abd08}. \textit{Hyperz} is a template fitting procedure and adopts a standard $\chi^2$ minimization algorithm:
\begin{equation}
\chi^2 = \sum_{i=1}^{N_{\rm filters}} \Bigl[ {F_{{\rm obs},i}-b \times F_{{\rm temp},i} \over \sigma_i} \Bigr] ^2 \label{eq.x2}
\end{equation}
where $F_{{\rm obs},i}$, $F_{{\rm temp},i}$ and $\sigma_{i}$ are the observed and template fluxes and their uncertainty in filter $i$, respectively, and $b$ is a normalization constant.

\textit{Hyperz} code is based on the fit of the overall shape of the spectra and on the detection of strong spectral features, such
as the 4000\,$\rm \AA$ break, Balmer break, Lyman decrement or
strong emission lines, { and} adopts a standard SED fitting method.

\textit{Hyperz} takes as its inputs the filter set and the photometric catalogue of galaxies which comprises magnitudes and photometric errors through the filters specified in the filter set. For a given filter set and galaxy catalogue, the relevant parameters introduced in the photo-$z$ calculation are: (i) the set of template spectra, which can be observed SEDs or built with spectral models or both (see this section){, if} spectral models { are used}, the type of SFR, the possible link between the age and the metallicity of the stellar population, and the choice of an initial mass function (IMF) are involved{ ;} (ii) the reddening law { (\textit{Hyperz} code provides 5 laws to choose);} (iii) flux decrements in the Lyman forest{ ;} (iv) the limiting magnitude in each filter{ ; and,} (v) the cosmological parameters $H_0, \Omega_{\rm M}$ and $\Omega_\Lambda$, and so on.

The descriptions about the filter set and the photometric catalogue of galaxies { \color{red} are presented in Section 3}. This section describes emphatically { the} template spectral library { (Sections 2.1 and 2.2) and the definition of models (see Section 2.3)}.

\subsection{Spectral synthesis models}
In the \textit{Hyperz} package observed SEDs and the spectral synthesis models of GISSEL98 (Galaxy Isochrone Synthesis Spectral Evolution Library, \citealt{bru93}) are provided. To check the effects of spectral synthesis models and binary interactions on the determinations of photo-$z$s and morphologies of galaxies, we also build the theoretical template SEDs by using the \citet[hereafter BC03]{bru03}, HPL07 and Yunnan models \citep{zha02,zha04a,zha04b,zha05}.

GISSEL98 and BC03 models were built by Bruzual \& Charlot in 1993 and 2003, respectively. In this work the GISSEL98 models with the IMF by \citet{mil79} and the BC03 models{ , which use} the Padova 1994 stellar evolutionary tracks and the \citet{cha03} IMF{ ,} are used because the choice of IMF has a negligible impact on the final results \citep{bol00}. Both models did not take binary interactions into account.

HPL07 models included the binary interactions of sdBs \citep{han02,han03} in populations. { The} Yunnan models were developed by Zhang and { her} colleagues, in { their} models the rapid single/binary evolution codes (\citealt*{hur00,hur02}) were used for the single/binary evolutionary tracks ({ the rapid binary evolution code includes those binary interactions studied before 2002}).
Both models presented the SEDs of populations without and with binary interactions, i.e., SSPs and BSPs.
To { distinguish them} we { name} the HPL07 { models for SSPs and BSPs HPL07-s and HPL07-b, name the} Yunnan models for SSPs and BSPs Yunnan-s and Yunnan-b models, respectively.
{ Detailed description of IMFs used by HPL07 and Yunnan models has presented in their related papers, we refer the interested reader to them.}

Moreover, the ages and age interval of these EPS models are different{. In Table}~\ref{Tab:ages} we give the initial and final ages [${\rm log}(t_i/{\rm yr})$, ${\rm log}(t_f/{\rm yr})$], the age interval [$\Delta {\rm log}(t/{\rm yr})$] within a given age range (in the brackets) for these EPS models.
In Fig.~\ref{Fig:isedcom-ath} we give { a} comparison { in the SED of} a solar-metallicity population (normalized to 1${\rm M_\odot}$) among BC03, HPL07 and Yunnan models at ${\rm log} (t/{\rm yr})=8.85$. From it we see that { the UV spectra} of SSP differs significantly from that of BSP for both HPL07 and Yunnan models, and the difference in the SED among BC03, HPL07-s and Yunnan-s models is also large{ . Therefore,} it is necessary to discuss the effects of spectral synthesis models, UV photometry and binary interactions on the determinations of { photo-$z$s and morphological types of galaxies}.

\subsection{Theoretical template SED library}
The template spectral library should comprise the SEDs with
different ages and spectral types of galaxies. Spectral synthesis
models only provide the SEDs of populations without any SFR at different ages{ . T}herefore, at a given age we need to generate the SEDs { of galaxies} with different galaxy types by means of spectral synthesis models. In this study we include Elliptical (E), Lenticular (S0), Spiral from Sa to Sd and Irregular (Irr) types.

Studies found that the observed properties of local field galaxies
with the different galaxy types can be roughly matched by
population with different SFRs.
Therefore, using the BC03 software package we build 8 SFRs { (}corresponding to the types from E to Irr{)}: a delta burst, a const star-formation system, and six $\mu$-models with characteristic time-decays (exponentially decreasing SFR, $\tau =1, 2, 3, 5, 10, 15$, and 30\,Gyr, see Eq.~\ref{eq.sfr}) chosen to match the sequence of colours from E-S0 to Sd \citep{bol00}:
\begin{equation}
$$\psi(t) = [1 + \varepsilon M_{\rm PG} (t)] \tau ^{-1} {\rm exp}(-t/\tau)$$
\label{eq.sfr}
\end{equation}
where $\tau$ is the e-folding timescale, $M_{\rm PG} (t)$ =
$[1-{\rm exp}(-t/\tau)] - M_{\rm stars} - M_{\rm remnants}$ is the
mass of gas that has been processed into stars and { then returned} to the interstellar medium (ISM) { due to stellar evolution} at time $t$, $M_{\rm stars}$ and $M_{\rm
remnants}$ are the masses of stars and remnants at $t$,
$\varepsilon$ denotes the fraction of $M_{\rm {PG}} (t)$ that can
be recycled into new star formation.

By convolving these 8 SFRs with the SEDs of populations we
generate 8 spectral families for a given spectral synthesis
models. Each family  { (corresponding to a certain SFR)} includes the SEDs at different ages.

In this study the number of ages (221, see {
Table}~\ref{Tab:ages}) is reduced to 51 for each spectral family
built with the GISSEL98 and BC03 models (refer to \citealt{bol00}
for details), while it remains unchanged for { the
family} built with the HPL07 and Yunnan models because the
\textit{Hyperz} code does not interpolate on the template grids
and the template set must be densely populated \citep{mar05}.

\subsection{Definition of Models}
The observed spectra, provided by the \textit{Hyperz} package, include 4 mean spectra of local E-, Sbc-, Scd- and Irr-type { galaxies} from \citet[][hereafter CWW, CWW-E, CWW-Sbc, CWW-Scd and CWW-Irr]{col80}{ . These spectra} extend from 1,400 to $10,000\,{\rm \AA}$ originally and { have been} extended to the UV and near-IR regions by means of GISSEL98 spectra with parameters (SFR and age) selected to match the observed spectra at $z=0$ \citep{bol00}.

In all of the following computations, the template SED library is constituted by 8 theoretical spectral families (correspond to Burst, E, S0-Sd and Irr types) generated by spectral synthesis models and 4 observed galaxy spectra from CWW.
{ For the sake of} clarity { 6 models are defined}, each model differs from the other by changing the set of theoretical template SEDs (8 spectral families) in the \textit{Hyperz} code: Models A/B/C/D use the theoretical template SEDs which were built with the GISSEL98/BC03/HPL07-s/Yunnan-s models (i.e., the models for SSPs), respectively, Models C2/D2 with the HPL07-b/Yunnan-b models (i.e., the models for BSPs).
All the other relevant parameters introduced in the computations are fixed except the set of filters and the catalogue of galaxies. The description of parameters { adopted by us} is as follows{ .}
\begin{itemize}
\item The combination of the CWW set of empirical SEDs with theoretical template spectra is used because the CWW set produces slightly better results than GISSEL98 (due to a Lyman blanketing effect) in the high redshift domain{ ,} and GISSEL98 models { produce} more accurate results than CWW templates alone in the low redshift domain.
    All 8 theoretical spectral families are used although there is no sensible gain when the set of 5 SFRs (1 delta burst, 3 { $\mu$-decaying}, 1 constant star-formation system, \citealt{bol00}) is used.
    Moreover, during the construction of theoretical template SEDs{, we (i) assume that} gas does not { get} recycled into new star formation once it { has been} processed into stars (i.e., $\varepsilon = 0${ , see Eq.~\ref{eq.sfr}}), { (ii) neglect} the attenuation by dust, { (iii) only use} solar-metallicity EPS models because \citet{bol00} found there is only a slight improvement on the accuracy of photo-$z$ at $z \la 1.5$ when several different metallicities are used together and { found that} including different metallicities does not affect the high redshift determinations, { and, (iv) only use} a certain IMF for each set of EPS models because that the photo-$z$ estimates are approximatively the same, whatever the IMF used \citep{bol00}.

\item { The} reddening law of \citet{cal00} is used to de-red the data because most of the fits to the galaxies from the Hubble Deep Field (HDF) { by} using the Calzetti's reddening law produce better $\chi^2$ values than { by} using the other attenuation laws provided by the \textit{Hyperz} package \citep{bol00}. In addition, in the computations $A_V$ [$\equiv R_V \cdot E(B-V)$] ranges from 0. to 1.2 in steps of 0.2.

\item The default values of flux decrements in the Lyman forest are used although { their variation} can { affect} the determination of photo-$z$, for example, decreasing them can induce { an} overestimate of photo-$z$.

\item { In} the case of non-detection in filter $i$ { (characterized by} the $i$-passband magnitude $m(i)=99$ { in the catalogue of galaxies), the contribution of the $i$-th filter} is also taken into account in the $\chi^2$ calculation { by assuming} the flux in the $i-$passband { is 0} { and the flux error equals to that deduced from the limiting magnitude ($\Delta F_{\rm obs} = \Delta F_{\rm lim}$).} This assumption is the usual setting when one is dealing with a relatively deep survey in the considered filter \citep{bol00}. The limiting magnitudes in the $F_{\rm _{UV}}N_{\rm _{UV}}ugriz$ passbands are set to 29, 29, 29, 30, 30, 30 and 30{ \,mag}, respectively.

\item { The} minimum of photometric error $\Delta m_{\rm min}$ is set to 0.05 because there is no significant gain for { $\Delta m_{\rm min} \le 0.05$} \citep[about 5\% accuracy, ][]{bol00}. When the input error from the catalogue is less than this value, { $\Delta m_{\rm min}$} will replace the input error to avoid too small and non-realistic photometric errors.

\item { The set of cosmological parameters $(\Omega_{\rm M}, \Omega_\Lambda, H_0) = (0.3, 0.7, 70)$ is used. The variation in this set of parameters [from $(1.0, 0.0, 50)$ to $(0.3, 0.7, 50)$] only affects $\delta_z$ by less than 1\% \citep{bol00}, here $H_0$ in units of ${\rm km\,s^{-1}\,Mpc^{-1}}$ and $\delta_z \equiv \sigma/(1+ \langle z \rangle { )}$, $\sigma = \sqrt{\Sigma_{i=1}^{N} |z_{\rm phot,i}-z_{\rm spec,i}|^2/(N-1)}$. }

\item { The} redshift ranges from 0 to 7 in a step of 0.02 instead of 0.05 (default value) although the primary $z$-step between 0.1 and 0.05 does not  significantly affect the results \citep{bol00}.
\end{itemize}

\section{Galaxy Sample}
\begin{table}
\centering
\caption{Characteristics of filters: the effective wavelength $\lambda_{\rm eff}$ and the surface of the normalized response function { (width)}. Note the superscript in the first column denotes the value of airmass.}
\begin{tabular}{lrr}
\hline
  filters     & $\lambda_{\rm eff}$[\AA] & width [\AA]  \\
\hline
 $F_{\rm _{UV}}$ &  1539.781  &   252.031  \\
 $N_{\rm _{UV}}$ &  2313.893  &   729.648  \\
 $u^0$        &  3523.768  &   587.708  \\
 $g^0$        &  4798.005  &  1327.291  \\
 $r^0$        &  6258.645  &  1333.650  \\
 $i^0$        &  7655.315  &  1345.910  \\
 $z^0$        &  8984.980  &  1193.529  \\
\hline
\end{tabular}
\label{Tab:flts}
\end{table}

\begin{figure}
\centering{
\includegraphics[bb=79 47 578 703,height=8.0cm,width=7.0cm,clip,angle=-90]{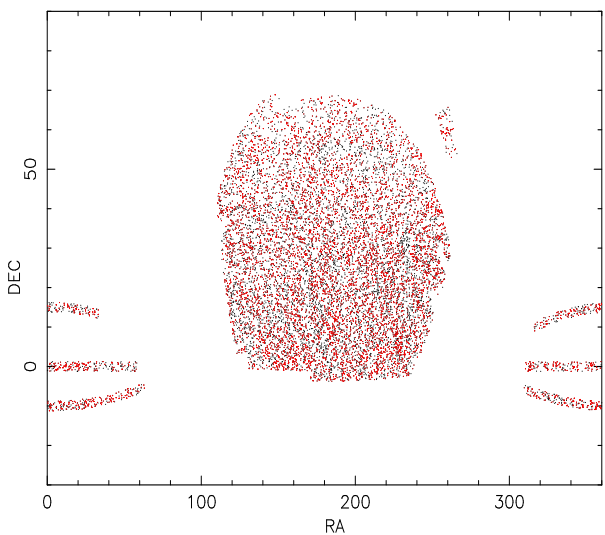}
}
\caption{The spatial distribution of galaxies in the spectroscopic sample (red { points}, 6,531 galaxies). For comparison those mismatched galaxies between SDSS DR7 and GALEX DR4 in the \textit{primate} spectroscopic sample{, i.e., those SDSS DR7 galaxies not included in GALEX DR4,} also are given (black points, 4,512 galaxies).}
\label{Fig:lowz-reg}
\end{figure}

\begin{figure}
\centering{
\includegraphics[bb=80 43 584 702,height=7.0cm,width=7.0cm,clip,angle=-90]{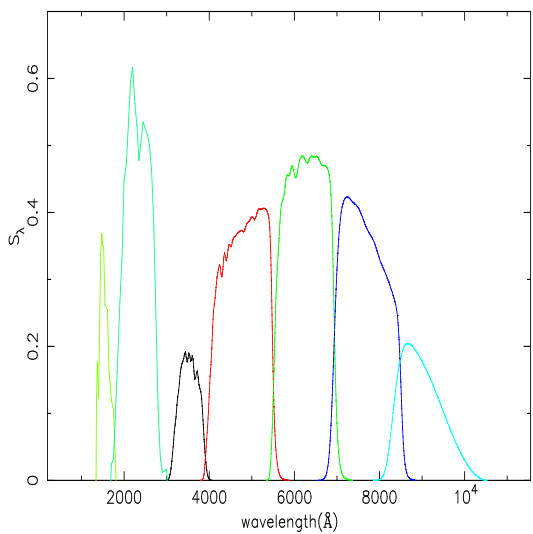}
}
\caption{The transmission curves $S_\lambda$ of $F_{\rm _{UV}}, N_{\rm _{UV}}$ filters in GALEX system and $ugriz^0$ (for a null airmass) filters in SDSS system.}
\label{Fig:flttran}
\end{figure}

To check the ability of recovering photo-$z$s and morphological types of galaxies, we need to collect the galaxies with known spectroscopic redshifts and morphological classifications, and the redshift coverage of galaxies { should be} as wide as possible.
But due to { the limit of observations} the galaxies with known classifications are relatively bright and their redshifts are low, two galaxy samples (termed spectroscopic and morphological) are used in this work, and they are used to study the ability of recovering photo-$z$s and galaxy types, respectively.

The galaxies in the spectroscopic sample { are selected} from the SDSS DR7{ . I}n order to ensure { that} the { selected} objects { can be used in the studies of Section 4, we use} the following criteria: the spectroscopic redshifts must be available (from \textit{SDSS.SpecObj}), and GALAXY must be defined as the object class by all of the following three methods: spectral classification (\textit{SDSS.SpecObj.specClass=2}), plPlugMap type (\textit{SDSS.SpecObj.objType=0}) and morphological classification (\textit{SDSS.Galaxy.type=3}).
Initially we obtain a sample of 773,126 galaxies from the SDSS DR7. To save the computational time we use one-seventy of this galaxy sample as the \textit{primitive} spectroscopic sample, which totally comprises 11,043 galaxies { and in which the galaxies are randomly selected}.

The galaxies in the morphological sample are from the catalogue of \citet{fuk07}, which contains 2,253 galaxies with \textit{Petrosian} magnitude $r_{\rm p}$ brighter than 16\,mag in the north equatorial stripe from the SDSS DR3. The morphological classification in the catalogue of \citet{fuk07} is obtained by visual inspection of images in the $g$ band. Removal the objects { mismatched} with the SDSS DR7 (the matching radius is 6\,arcsec) { produces} a initial morphological sample of 1,792 galaxies. Note that the redshifts of galaxies in this catalogue are low (less than 0.2).

\subsection{Matching the SDSS/DR7 sample with GALEX/DR4}
We adopt the matching method of \citet{agu05} and \citet{obr06} between the SDSS DR7 and the GALEX DR4, i.e., the matching radius between SDSS DR7 and GALEX DR4 is 6.0\,arcsec, the nearest neighbor is taken as a true associate if its distance is smaller than the matching radius{ . When} the nearest object has no simultaneous $F_{\rm _{UV}}$ and $N_{\rm _{UV}}$ detections, the nearer object (within the matching radius) with simultaneous $F_{\rm _{UV}}$ and $N_{\rm _{UV}}$ detections will be taken as its { associate} (361 galaxies for spectroscopic sample). If all objects within the matching radius have no simultaneous $F_{\rm _{UV}}$ and $N_{\rm _{UV}}$ detections, the missing magnitude of the nearest object will be replaced by the corresponding one of the nearer objects (within the matching radius).

At last, we obtain a final spectroscopic sample of 6,531 galaxies{ . I}n Fig.~\ref{Fig:lowz-reg} we give the spatial distribution of these matched (i.e., the spectroscopic sample) and mismatched galaxies in the \textit{primitive} spectroscopic sample, from it we see that the galaxies in the spectroscopic sample are indeed randomly distributed. Also we obtain a final morphological sample of 1,502 galaxies. The spectroscopic and morphological galaxy samples are used in the { studies of Sections} 4 and 5, respectively.

SDSS DR7 provides { the values of various} magnitudes (such as \textit{devMag, expMag, modelMag, petroMag, psfMag,} etc){ . W}e choose SDSS \textit{modelMag} (exactly, the shorthand alias for \textit{modelMag}) of galaxies { since t}he SDSS \textit{C-model} magnitude (based on de Vaucouleurs and exponential fitting) is an adequate proxy to use as a universal magnitude for all types of objects { and} agrees with \textit{Petrosian} magnitude excellently for galaxies{. Moreover, the SDSS \textit{C-model} magnitude} is a reliable estimate of galaxy flux { and} can account for the effects of local seeing and thus is less dependent on local seeing variations (www.sdss.org/DR7/algorithms/photometry.html).

\subsection{Filters}
The filters, involved in the computations, include those equipped in the SDSS system and those in the GALEX system. The transmission curves ($S_{\rm \lambda}$) of these filters are presented in Fig.~\ref{Fig:flttran}. The data about the SDSS filters for a null airmass ($ugriz^0$) are from the BC03 package. The characteristics of these filters (the effective wavelength $\lambda_{\rm eff}$ and the surface of the normalized response function [width]) are presented in { Table}~\ref{Tab:flts}.

\section{UV photometry, Binary interactions and population synthesis models on photo-$z$ determinations}

\begin{figure}
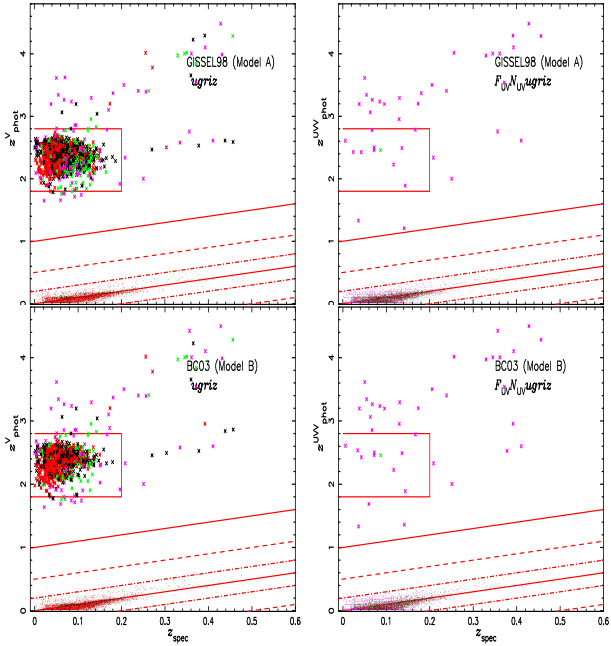

\centering{
\includegraphics[bb=81 35 530 721,height=4.0cm,width=4.0cm,clip,angle=-90]{4zspec-zphot-V-GISSEL98.ps}
\includegraphics[bb=81 35 530 721,height=4.0cm,width=4.0cm,clip,angle=-90]{4zspec-zphot-UVV-GISSEL98.ps}
\includegraphics[bb=81 26 580 712,height=4.0cm,width=4.5cm,clip,angle=-90]{4zspec-zphot-V-bc03.ps}
\includegraphics[bb=81 26 580 712,height=4.0cm,width=4.5cm,clip,angle=-90]{4zspec-zphot-UVV-bc03.ps}
}
\caption{Comparisons between spectroscopic ($z_{\rm spec}$) and photometric ($z_{\rm phot}$) redshifts for the spectroscopic galaxy sample (see { Section} 3.2). Left panels are for only using optical ($ugriz$) photometry. Right panels are for using the combination of UV with optical photometry ($F_{\rm _{UV}}N_{\rm _{UV}}ugriz$). Top and bottom panels are for Models A and B, respectively. 
In each panel dot-dashed, dashed and solid lines correspond to $|\Delta z|=0.2, 0.5$ and 1.0. Dots and crosses denote galaxies with $z_{\rm phot} \le 1.0$ and $z_{\rm phot} > 1.0$. Red, black, green and purple symbols denote galaxies with the photo-$z$ probability $P(\chi^2) > 99, 99 \ge P(\chi^2) > 90, 90 \ge
P(\chi^2) > 68, P(\chi^2) \le 68$ (see Eq.~\ref{eq.x2}),
respectively.}
\label{Fig:lowz-dia1}
\end{figure}

\begin{figure}
\centering{
\includegraphics[bb=81 35 530 721,height=4.0cm,width=4.0cm,clip,angle=-90]{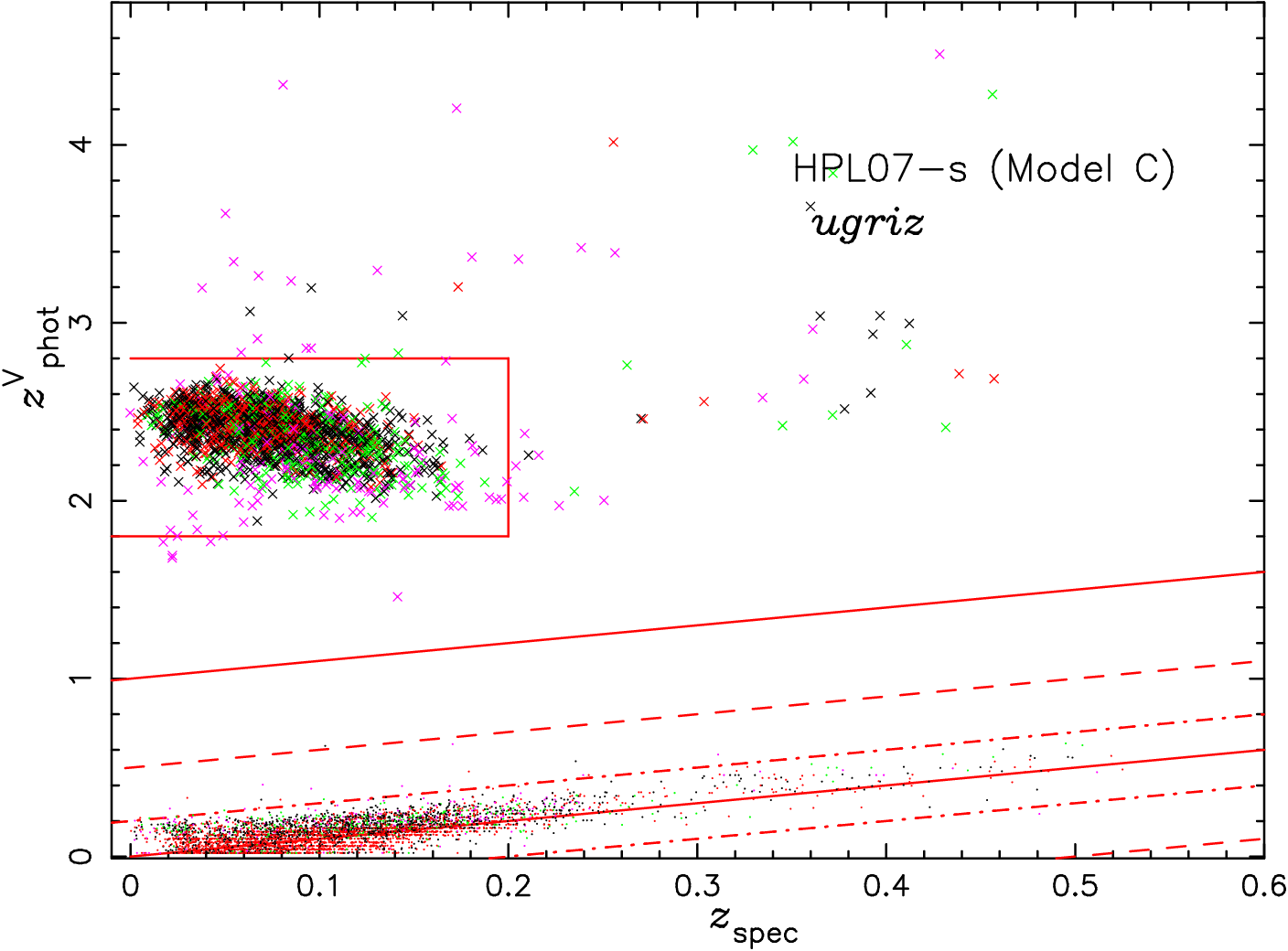}
\includegraphics[bb=81 35 530 721,height=4.0cm,width=4.0cm,clip,angle=-90]{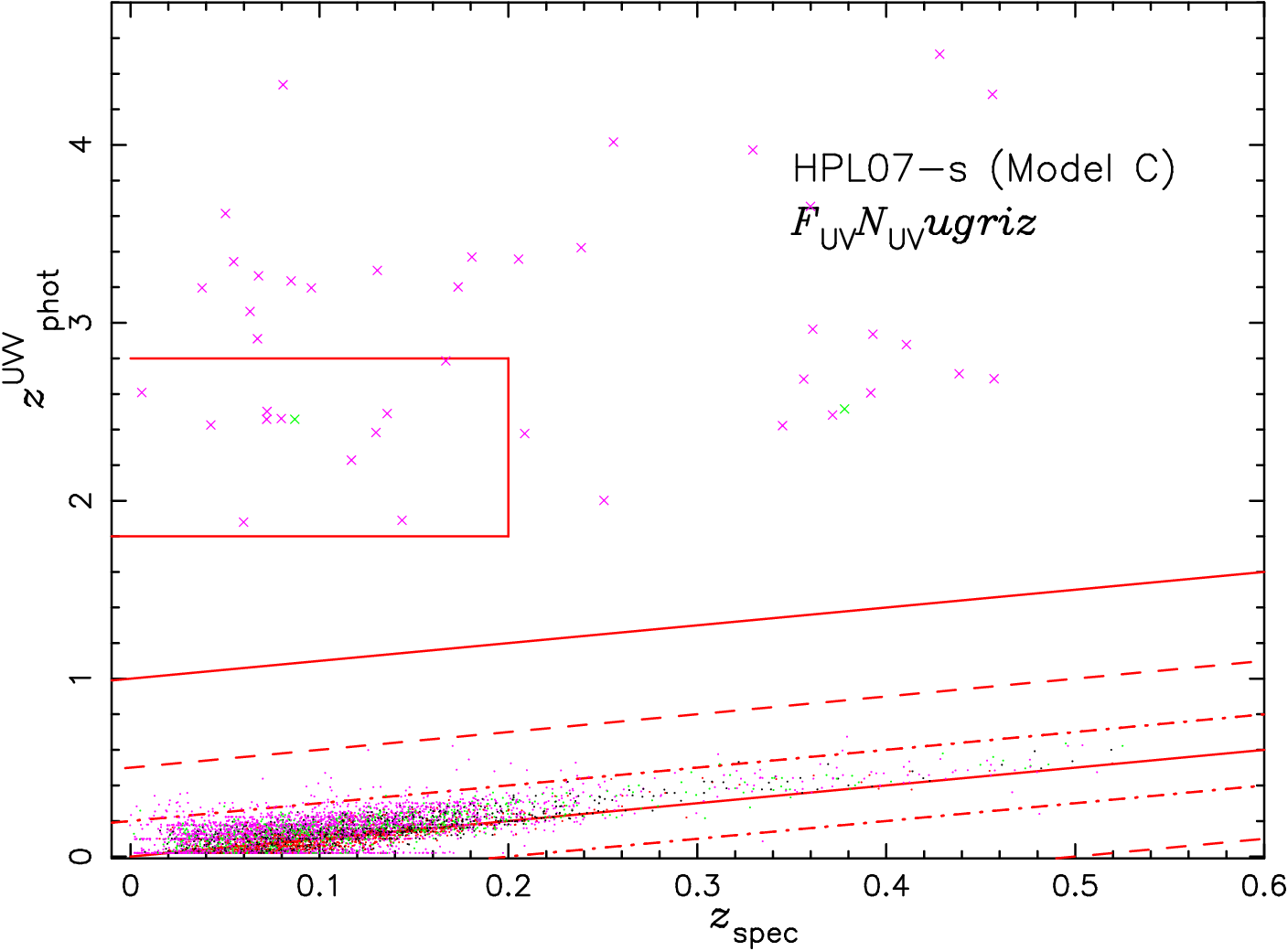}
\includegraphics[bb=81 26 580 712,height=4.0cm,width=4.5cm,clip,angle=-90]{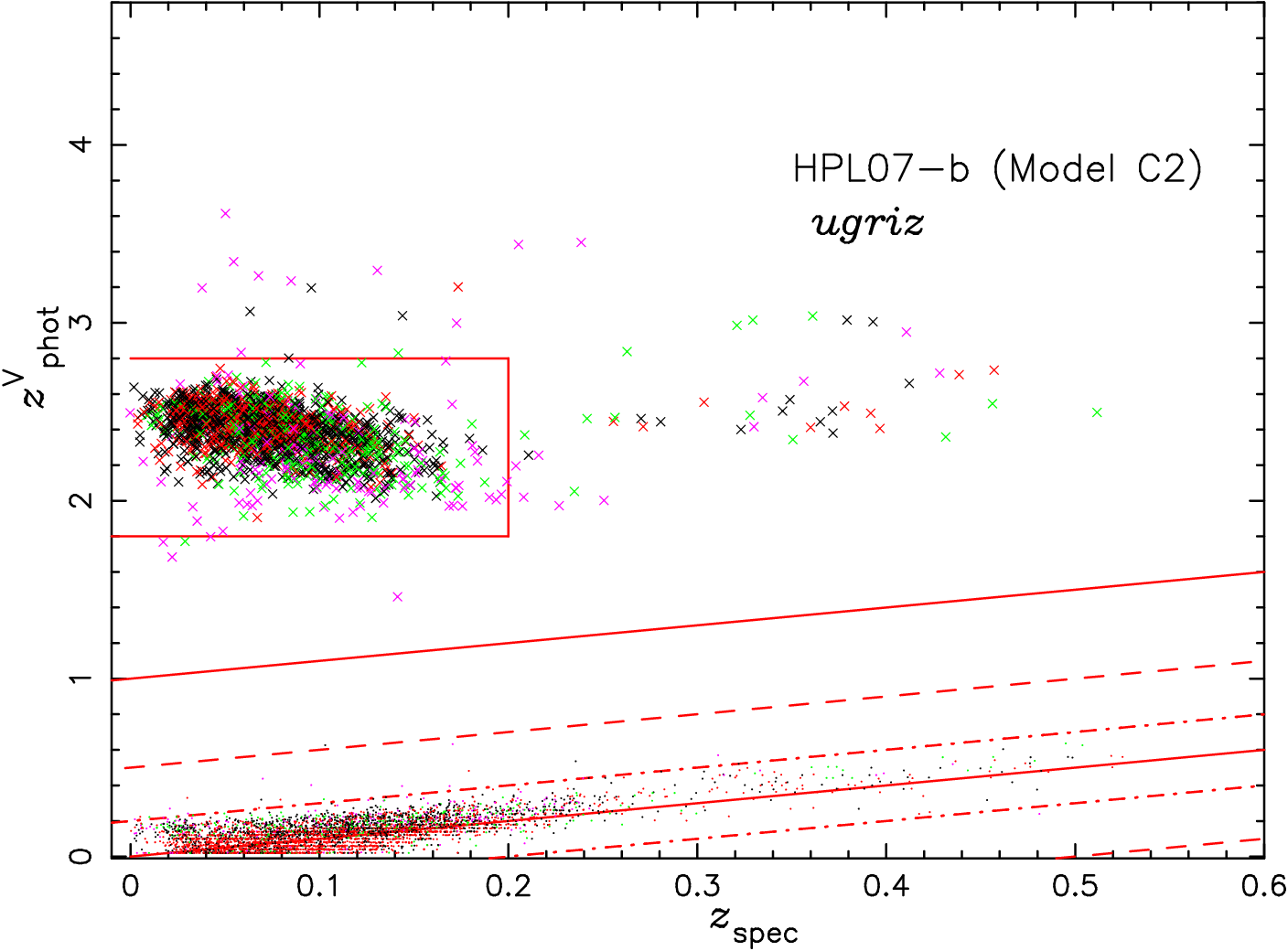}
\includegraphics[bb=81 26 580 712,height=4.0cm,width=4.5cm,clip,angle=-90]{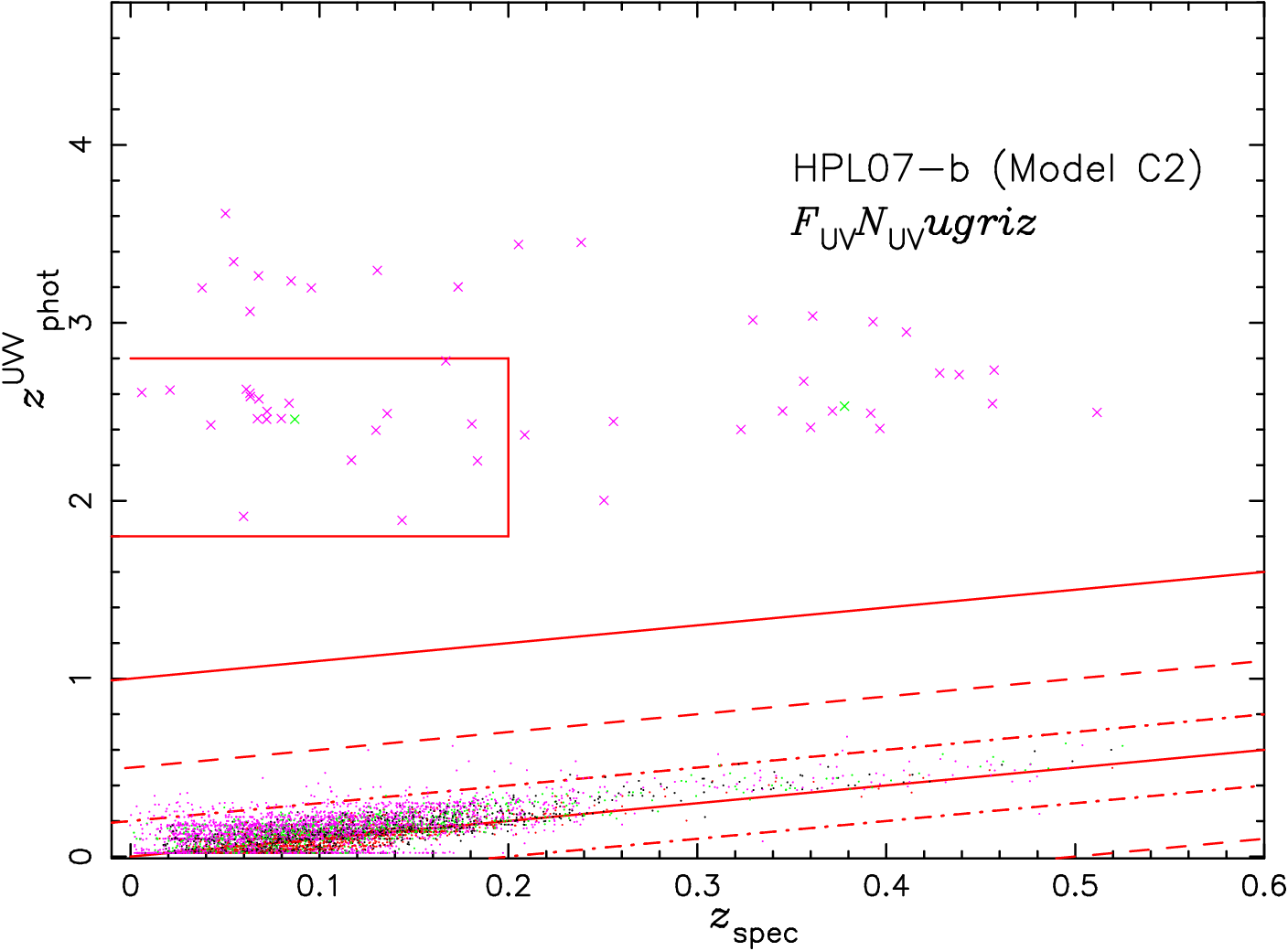}
}
\caption{Similar to Fig.~\ref{Fig:lowz-dia1}, but for Models C and C2 (top and bottom panels).}
\label{Fig:lowz-dia-hpl}
\end{figure}

\begin{figure}
\centering{
\includegraphics[bb=81 35 530 721,height=4.0cm,width=4.0cm,clip,angle=-90]{4zspec-zphot-V-ynss.ps}
\includegraphics[bb=81 35 530 721,height=4.0cm,width=4.0cm,clip,angle=-90]{4zspec-zphot-UVV-ynss.ps}
\includegraphics[bb=81 26 580 712,height=4.0cm,width=4.5cm,clip,angle=-90]{4zspec-zphot-V-ynbb.ps}
\includegraphics[bb=81 26 580 712,height=4.0cm,width=4.5cm,clip,angle=-90]{4zspec-zphot-UVV-ynbb.ps}
}
\caption{Similar to Fig.~\ref{Fig:lowz-dia1}, but for Models D and D2 (top and bottom panels).}
\label{Fig:lowz-dia-yn}
\end{figure}

\begin{figure}
\centering{
\includegraphics[bb=81 42 586 700,height=7.0cm,width=7.0cm,clip,angle=-90]{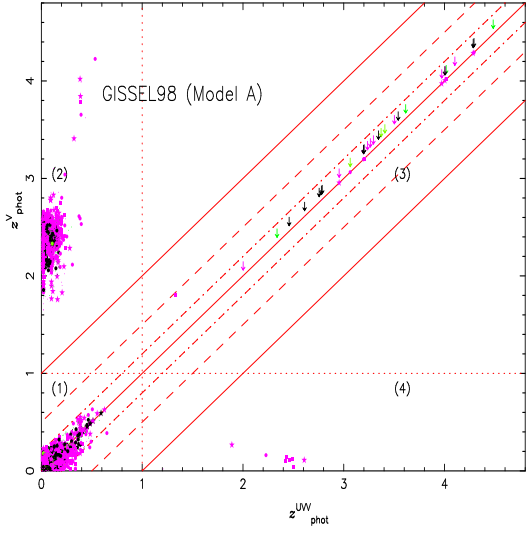}
}
\caption{Comparison between $z{\rm_{phot}^{V}}$ and $z{\rm_{phot}^{UVV}}$ { (based on Model A)} for the spectroscopic galaxy sample. The dot-dashed, dashed and { solid} lines have the same meaning as in Fig.~\ref{Fig:lowz-dia1}.
The numbers in this figure correspond to the spectroscopic subsample (see { Section} 4.1.2). Arrows point to the location of galaxies in the 3rd subsample for the sake of clarity. Different symbols (square{s}, circle{s}, star{s} and dot{s}) and colours (red, black, green and purple) denote galaxies with different $z_{\rm phot}^{\rm V}$ and $z_{\rm phot}^{\rm UVV}$ probability: $P(\chi^2) > 99, 99 \ge P(\chi^2) > 90, 90 \ge P(\chi^2) > 68, P(\chi^2) \le 68$ (see Eq.~\ref{eq.x2}), respectively. }
\label{Fig:zcom-psb-UVv}
\end{figure}

\begin{figure}
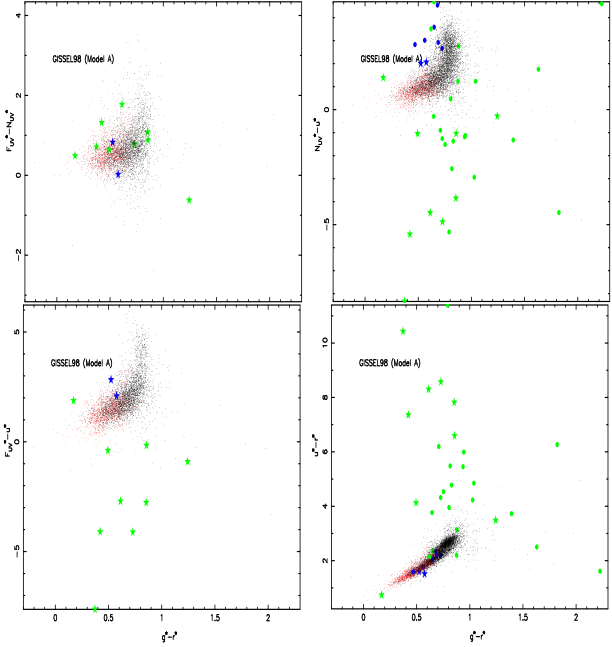

\centering{
\includegraphics[bb=81 49 531 706,height=4.0cm,width=4.0cm,clip,angle=-90]{4colrange-gr-FUVNUV-GISSEL98.ps}
\includegraphics[bb=81 49 531 706,height=4.0cm,width=4.0cm,clip,angle=-90]{4colrange-gr-NUVu-GISSEL98.ps}
\includegraphics[bb=81 44 580 701,height=4.0cm,width=4.5cm,clip,angle=-90]{4colrange-gr-FUVu-GISSEL98.ps}
\includegraphics[bb=81 44 580 701,height=4.0cm,width=4.5cm,clip,angle=-90]{4colrange-gr-ur-GISSEL98.ps}
}
\caption{The colour-colour diagrams for the spectroscopic galaxy sample. The galaxies in the 1-4 subsamples{ , which are divided based on Model A,} are in black, red, green and blue, respectively. Squares, circles and stars denote galaxies with $F_{\rm _{UV}}ugriz$, $N_{\rm _{UV}}ugriz$ and $F_{\rm _{UV}}N_{\rm UV}ugriz$ detections in the 3-4 subsamples, respectively.}
\label{Fig:colrange}
\end{figure}

\begin{figure}
\centering{
\includegraphics[bb=79 44 586 700,height=7.0cm,width=7.0cm,clip,angle=-90]{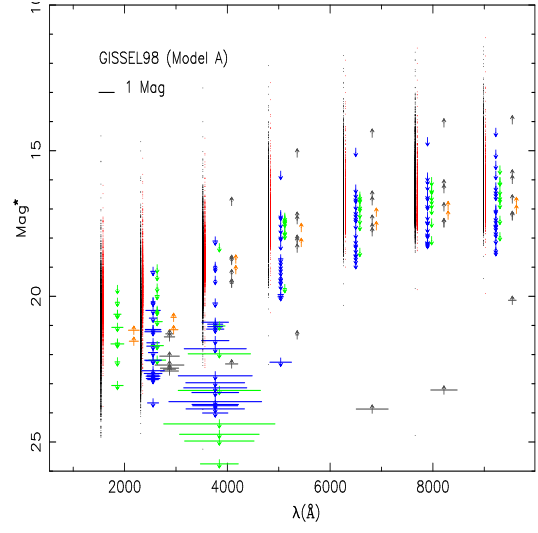}
}
\caption{Distribution of magnitudes (from left to right: $F_{\rm _{UV}}^\star,N_{\rm _{UV}}^\star, u^\star,g^\star,r^\star,i^\star,z^\star$) along the wavelength for the 4 spectroscopic galaxy subsamples, which are divided based on Model A. Black dots, red dots, downarrows and uparrows represent the galaxies in the 1-4 subsamples, respectively. The horizontal lines represent the corresponding magnitude-errors{ , $\Delta Mag$. The length of line located in the upper left corner denotes $\Delta Mag =1$.} The galaxies with $F_{\rm _{UV}}N_{\rm _{UV}}ugriz$ detections in the 3rd and 4th subsamples are in green and purple, and those with $N_{\rm _{UV}}ugriz$ detections are in blue and grey. { For the sake of} clarity, the galaxies in the 2-4 subsamples are shifted rightwards.}
\label{Fig:maglam}
\end{figure}

\begin{figure}
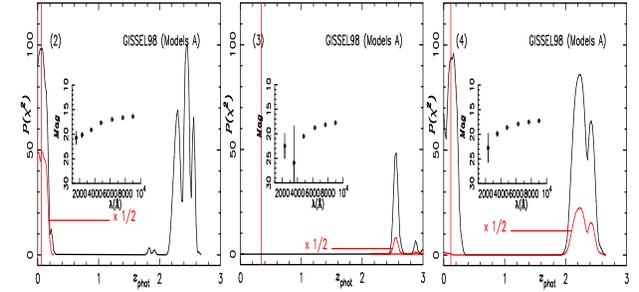

\centering{
\includegraphics[bb=75  34  589 705 ,height=2.7cm,width=4.0cm,clip,angle=-90]{px2-GISSEL98-2sub7-2.ps}\includegraphics[bb=75  34  589 705 ,height=2.7cm,width=4.0cm,clip,angle=-90]{px2-GISSEL98-3sub28-2.ps}\includegraphics[bb=75  34  589 705 ,height=2.7cm,width=4.0cm,clip,angle=-90]{px2-GISSEL98-4sub3-2.ps}
}
\caption{Probability { of photometric redshift} $P(\chi^2)$ (see Eq.~\ref{eq.x2}) as a function of redshift for an arbitrary galaxy in the 2nd (left panel), 3rd (middle panel) and 4th (right panel) subsamples based on Model A{ . In each panel} black and red curves (the probability has been reduced by half for the sake of clarity) are for using $ugriz$ and $F_{\rm _{UV}}N_{\rm _{UV}}ugriz$ photometry. Red vertical line { corresponds to} the spectroscopic value. Also in each panel the magnitudes and their uncertainties, which are multiplied by 5, are plotted for the corresponding galaxy.}
\label{Fig:px2}
\end{figure}


\begin{figure}
\centering{
\includegraphics[bb=80 124 541 659,height=7.5cm,width= 7.5cm,clip,angle=-90]{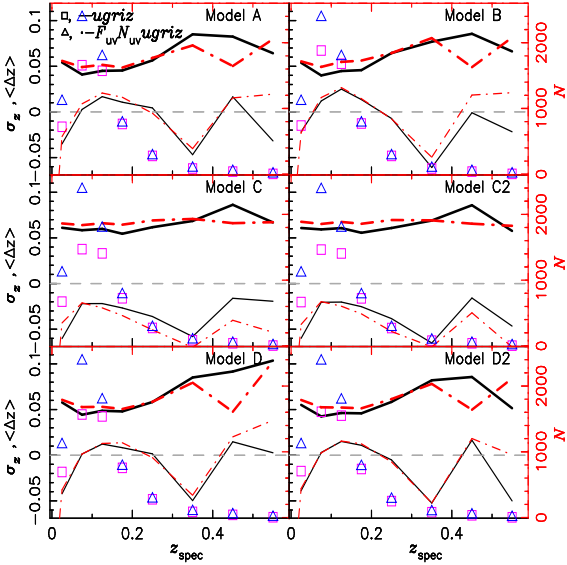}
}
\caption{ the number of non-catastrophic identifications $N$ (squares$-ugriz$; triangles$-F_{\rm _{UV}}N_{\rm _{UV}}ugriz$), mean deviation $\langle \Delta z \rangle$ (thin lines) and standard deviation $\sigma_z$ (thick lines) [solid line$-ugriz$; dot-dashed line$-F_{\rm _{UV}}N_{\rm _{UV}}ugriz$] as a function of $z_{\rm spec}$ in removing catastrophic identifications for Models A-D2.}
\label{Fig:sigma-z}
\end{figure}

\begin{figure}
\centering{
\includegraphics[bb=80 124 541 659,height=7.5cm,width= 7.5cm,clip,angle=-90]{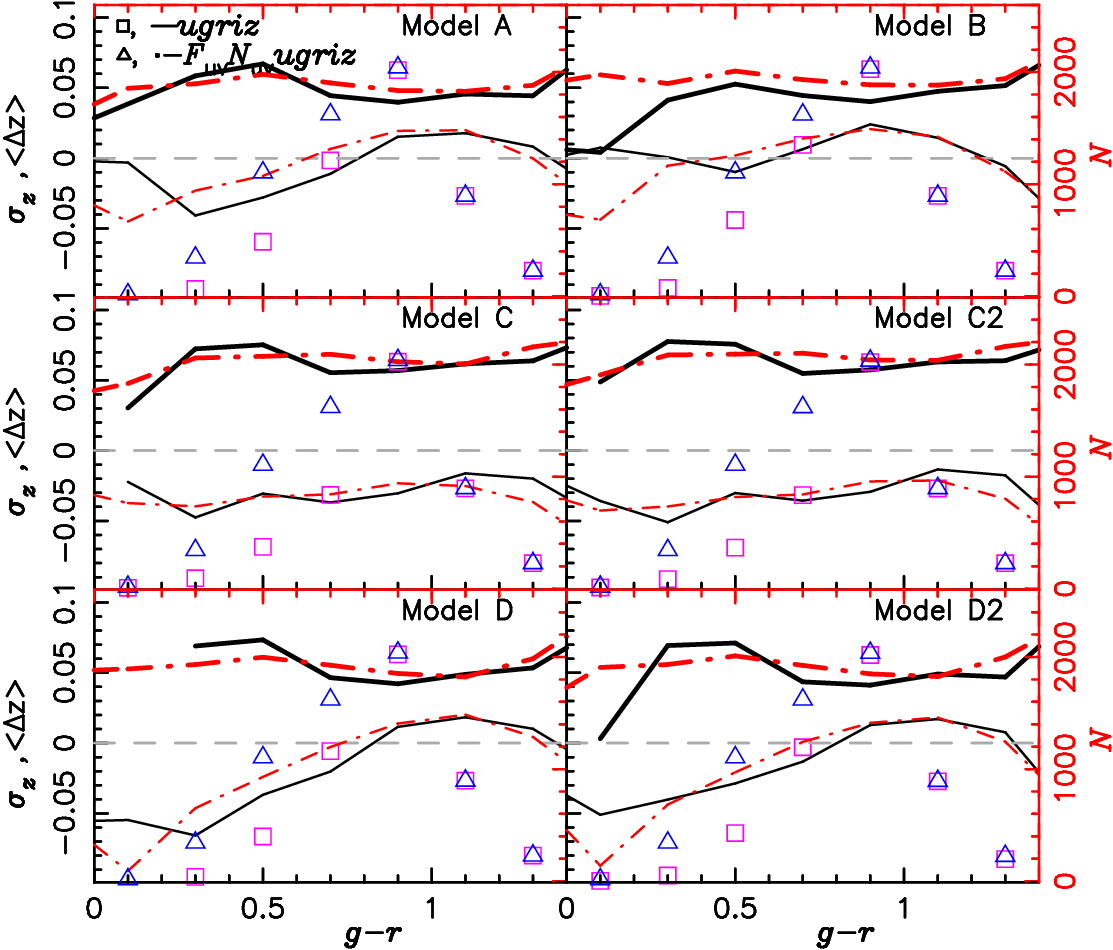}
}
\caption{ Similar to Fig.~\ref{Fig:sigma-z}, but as a function of $g-r$ colour.}
\label{Fig:sigma-gr}
\end{figure}

\begin{figure}
\centering{
\includegraphics[bb=80 124 541 659,height=7.5cm,width= 7.5cm,clip,angle=-90]{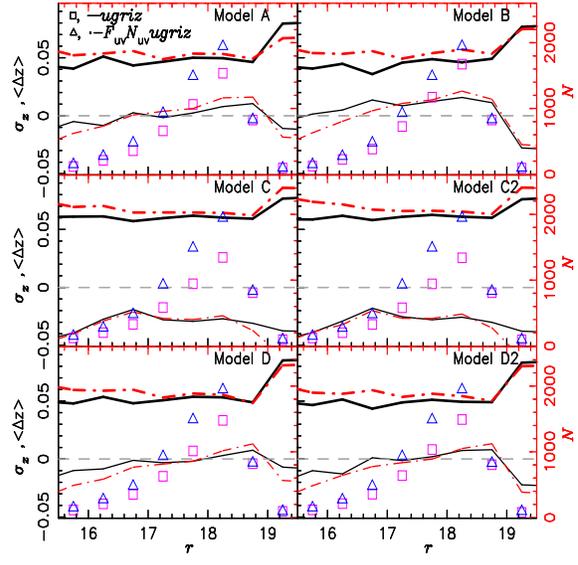}}
\caption{ Similar to Fig.~\ref{Fig:sigma-z}, but as a function of $r-$magnitude.}
\label{Fig:sigma-r}
\end{figure}

\begin{figure}
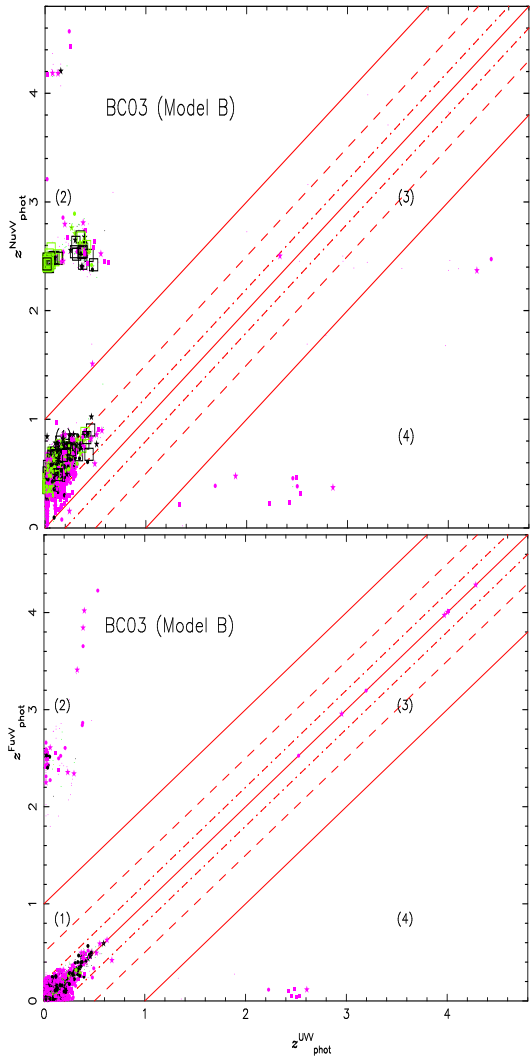

\centering{
\includegraphics[bb=79 48 532 706,height=7.0cm,width=7.0cm,clip,angle=-90]{4zphot-UVNUV-bc03.ps}
\includegraphics[bb=79 45 586 703,height=7.0cm,width=7.0cm,clip,angle=-90]{4zphot-UVFUV-bc03.ps}
}
\caption{{ Comparison} between $z{\rm_{phot}^{N_{\rm _{UV}}V}}$ and $z{\rm_{phot}^{UVV}}$ (top panel{, based on Model B}) and comparison between $z{\rm_{phot}^{F_{\rm _{UV}}V}}$ and $z{\rm_{phot}^{UVV}}$ (bottom panel{, based on Model B}) for the spectroscopic galaxy sample. The lines, symbols and colours have the same meaning as in Fig.~\ref{Fig:zcom-psb-UVv}, but the y-axis { represents} $z{\rm_{phot}^{N_{\rm _{UV}}V}}$ or $z{\rm_{phot}^{F_{\rm _{UV}}V}}$, respectively.}
\label{Fig:lowz-dia-bc03-NUV}
\end{figure}

\begin{figure}
\centering{
\includegraphics[bb=69 48 586 720,height=8.0cm,width=7.0cm,clip,angle=-90]{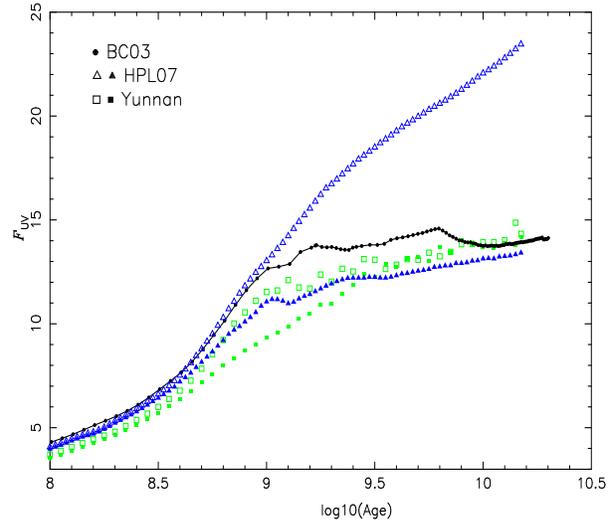}
}
\caption{$F_{\rm _{UV}}$ magnitude of a solar-metallicity population (normalized to 1\,${\rm M}_\odot$) for HPL07 (triangles) and Yunnan (squares) models. Open and solid symbols { represent} SSP and BSP, respectively. Also shown are the results of BC03 ({ solid black line+}circles).}
\label{Fig:FUVcom}
\end{figure}

\begin{figure}
\centering{
\includegraphics[bb=69 48 586 720,height=8.0cm,width=7.0cm,clip,angle=-90]{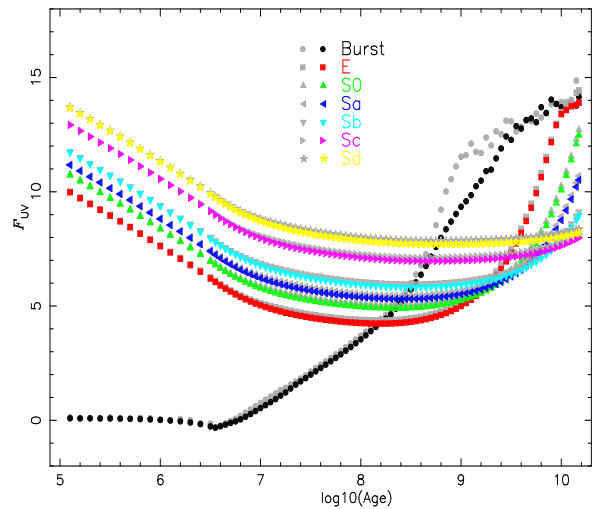}
}
\caption{$F_{\rm _{UV}}$ magnitude of a solar-metallicity galaxy (1\,${\rm M}_\odot$) at $z=0$.
{ Solid circles, squares, up-triangles, left-triangles, right-triangles, down-triangles and stars symbols are for Burst, E, S0, Sa, Sb, Sc and Sd galaxies. The gray and coloured (black, red, green, blue, cyan, magenta and yellow correspond to Burst, E, S0, Sa-Sd types, in turn) symbols} refer to using Yunnan-s and Yunnan-b models, respectively.
}
\label{Fig:FUVcspcom}
\end{figure}


\begin{figure}
\centering{
\includegraphics[bb=79 86 572 597,height=6.50cm,width= 7.00cm,clip,angle=-90]{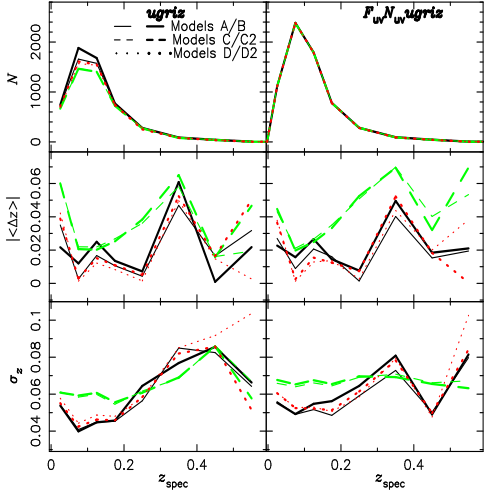}
}
\caption{ Evolutions of $N$ (top panels), $|\langle \Delta z \rangle|$ (middle panels) and $\sigma_z$ (bottom panels) as a function of $z_{\rm spec}$ for Models A-D2 in removing catastrophic identifications. Left an right panels are for using $ugriz$ and $F_{\rm _{UV}}N_{\rm _{UV}}ugriz$ photometry, respectively.
}
\label{Fig:sigma-z-models}
\end{figure}

\begin{figure}
\centering{
\includegraphics[bb=79 86 572 597,height=6.50cm,width= 7.00cm,clip,angle=-90]{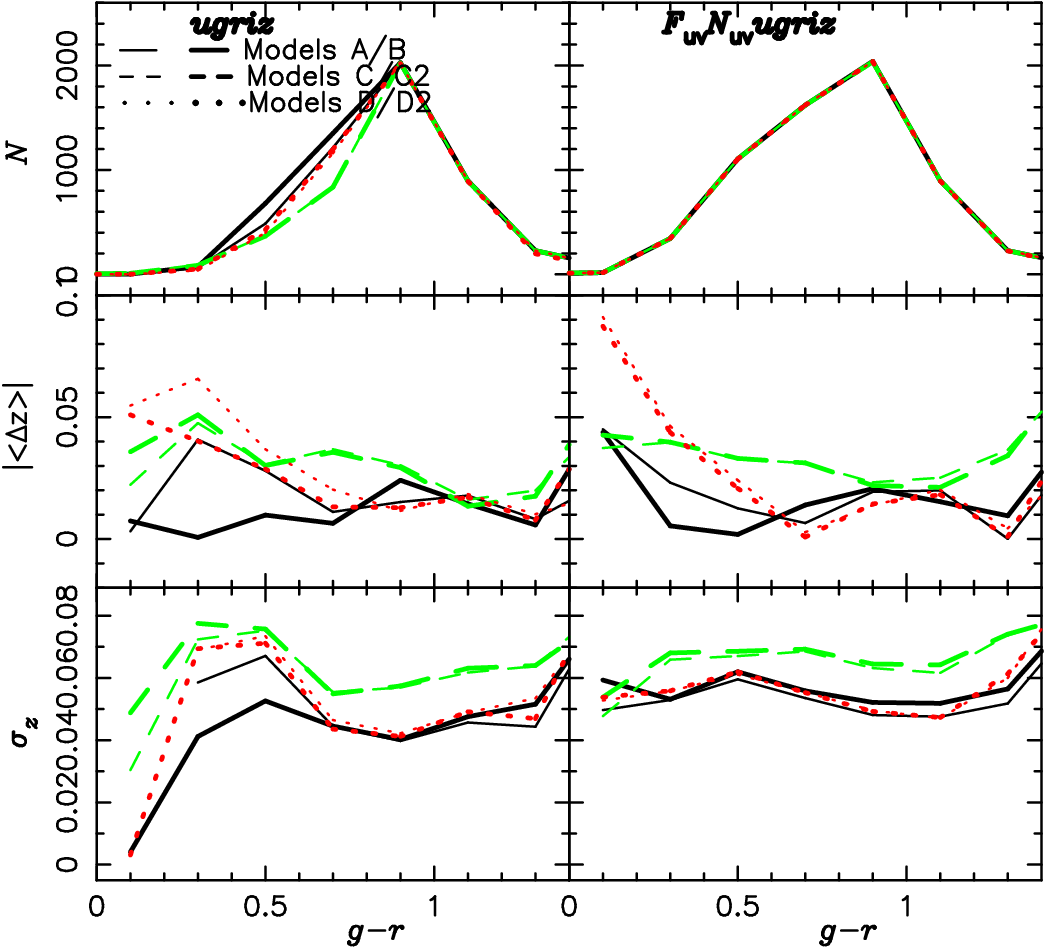}
}
\caption{ Similar to Fig.~\ref{Fig:sigma-z-models}, but a function of $g-r$ colour.} \label{Fig:sigma-gr-models}
\end{figure}

\begin{figure}
\centering{
\includegraphics[bb=79 86 572 597,height=6.50cm,width= 7.00cm,clip,angle=-90]{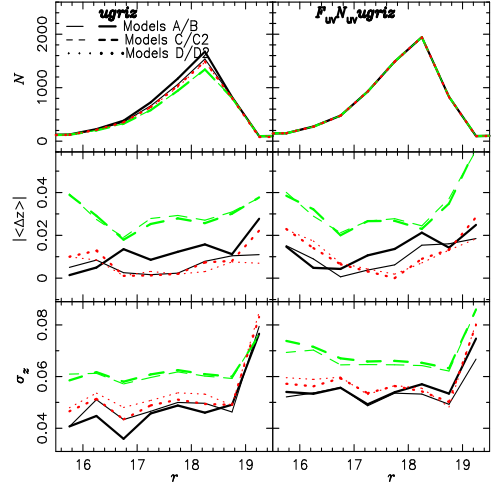}}
\caption{ Similar to Fig.~\ref{Fig:sigma-z-models}, but a function of $r-$magnitude.}
\label{Fig:sigma-r-models}
\end{figure}

\begin{figure}
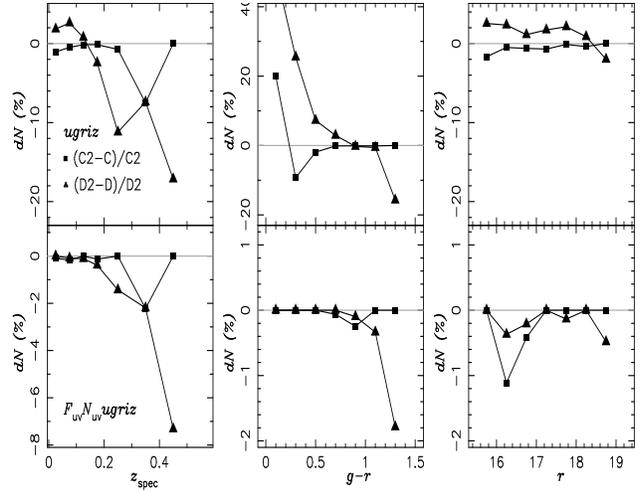

\centering{
\includegraphics[bb=53 83 547 361,height=2.70cm,width= 6.50cm,clip,angle=-90]{dn-z2}
\includegraphics[bb=53 83 547 361,height=2.70cm,width= 6.50cm,clip,angle=-90]{dn-gr2}
\includegraphics[bb=53 83 547 361,height=2.70cm,width= 6.50cm,clip,angle=-90]{dn-r2}
}
\caption{ The percent difference in the number of non-catastrophic identifications ${\rm d}N$ [$\equiv (N_{\rm x2}-N_{\rm x})/N_{\rm x2}$, x=C or D] between Models C and C2 (squares) and between Models D and D2 (triangles) as a function of $z_{\rm spec}$ (Left panels), $g-r$ colour (middle panels) and $r-$magnitude (right panels). Upper and bottom panels are for using $ugriz$ and $F_{\rm _{UV}}N_{\rm _{UV}}ugriz$ photometry, respectively.}
\label{Fig:dn-bi}
\end{figure}

In this section we will use the \textit{Hyperz} code to obtain photo-$z$ estimates of galaxies in the spectroscopic sample, and discuss the effects of UV photometry{ ,} binary interactions { and population synthesis models} on photo-$z$ determinations.

All parameters adopted by us in the \textit{Hyperz} procedure (including the set of template spectra, reddening law and so on) have described in { Section} 2.3. In Appendix A we present the influence of these parameters on photo-$z$ estimates of galaxies in the spectroscopic sample.

\subsection{UV photometry on photo-$z$ determinations}
\subsubsection{The effect of UV photometry}
We obtain photo-$z$ estimates ($z{\rm_{phot}^V}$) of galaxies in the spectroscopic sample by only using SDSS $ugriz$ data and those ($z{\rm_{phot}^{UVV}}$) by using the full set of $F_{\rm _{UV}}N_{\rm _{UV}}ugriz$ photometry for Models A-D2 (see { Section} 2.3).

In the left panels of Figs.~\ref{Fig:lowz-dia1}-\ref{Fig:lowz-dia-yn}, we give the comparisons between SDSS spectroscopic ($z{\rm_{spec}}$) and photometric ($z{\rm_{phot}^V}$) redshifts for Models A-D2, respectively.
From them we see that for most of galaxies ($\sim$71.3\% for Model C) $z{\rm_{phot}^V}$ is less than 1.0 and the difference between spectroscopic and photometric redshifts $|\Delta z^{\rm V}|$ ($\equiv |z_{\rm spec}-z_{\rm phot}^{\rm V}|$) is less than 0.2{, and} for part of galaxies ($\sim$27.4\% for Model C) $z{\rm_{phot}^V}$ ranges from 1.8 to 2.8 but { $z_{\rm spec}$ is} less than 0.2 (these galaxies are located within the { corresponding} box of the left panels in Figs.~\ref{Fig:lowz-dia1}-\ref{Fig:lowz-dia-yn}).

In the right panels of Figs.~\ref{Fig:lowz-dia1}-\ref{Fig:lowz-dia-yn} we give the comparisons between $z{\rm_{spec}}$ and $z{\rm_{phot}^{\rm UVV}}$ for Models A-D2.
By comparing with the corresponding left-hand panel, we find that the number of galaxies with $z_{\rm phot}^{\rm UVV} < 1.0$ and $|\Delta z^{\rm UVV}| (\equiv |z_{\rm spec} - z_{\rm phot}^{\rm UVV}|) < 0.2$ { is largely increased} ($\sim 98.0\%$ for Model C), and that of galaxies with $1.8 \le z{\rm_{phot}^{\rm UVV}} \le 2.8$ and $z{\rm_{spec}} \le 0.2$ { is significantly decreased}  ($\sim 0.18\%$ for Model C).
That is to say, the inclusion of UV photometry { would make} the difference between photometric and spectroscopic redshifts ($|\Delta z|$) smaller (from 1.8-2.8 to less than 0.2) for these galaxies. When we only use optical data to obtain the { photo-$z$ estimates} of these galaxies, we must be cautious. Then what are the characteristics of these galaxies?

\subsubsection{which galaxies are affected?}
According to the results about $z{\rm_{phot}^V}$ and $z{\rm_{phot}^{UVV}}$, we divide the spectroscopic galaxy sample into 4 subsamples for each model, the galaxies in the 1-4 subsamples satisfy the following conditions in turn:
\begin{enumerate}
\item both $z{\rm_{phot}^V}$ and $z{\rm_{phot}^{UVV}}$ are less than 1.0, and the differences, $|\Delta z^{\rm V}|$ and $|\Delta z^{\rm UVV}|$, are less than 0.2;
\item $z{\rm_{phot}^V} \ge 1.0$ and $z{\rm_{phot}^{UVV}} \le 1.0$;
\item both $z{\rm_{phot}^V}$ and $z{\rm_{phot}^{UVV}}$ are greater than 1.0;
\item $z{\rm_{phot}^V} \le 1.0$ and $z{\rm_{phot}^{UVV}} \ge 1.0$.
\end{enumerate}
This division means that the photo-$z$ estimates { of galaxies in the 1st subsample} could be obtained { reasonably} { in the case of only using optical $ugriz$ data}, the { photo-$z$ errors} of galaxies in the 2nd subsample { could be decreased by the combination of UV with optical data, the photo-$z$ estimates of galaxies in the 3rd subsample could not be obtained reasonably even if UV data is combined, and the photo-$z$ estimates of galaxies in the 4th subsample would be erroneously identified if UV photometry is included.}
In Fig.~\ref{Fig:zcom-psb-UVv} we give { a} comparison between $z{\rm_{phot}^V}$ and $z{\rm_{phot}^{UVV}}$ for Model A, the galaxies in the 1-4 subsamples occupy the lower left (except 80 galaxies, for which $0.2 < \Delta z <0.5$), upper left, upper right and lower right regions, respectively.

To investigate the properties of galaxies in each subsample, according to the results of Model A, in Fig.~\ref{Fig:colrange} we present the distribution of galaxies in the diagrams of ($F_{\rm _{UV}}^\star-N_{\rm _{UV}}^\star$).vs.($g^\star - r^\star$), ($F_{\rm _{UV}}^\star-u^\star$).vs.($g^\star - r^\star$), ($N_{\rm _{UV}}^\star-u^\star$).vs.($g^\star - r^\star$) and ($u^\star-r^\star$).vs.($g^\star - r^\star$),
in Fig.~\ref{Fig:maglam} present the magnitude ($F_{\rm _{UV}}^\star, N_{\rm _{UV}}^\star,u^\star,g^\star,r^\star,i^\star$ and $z^\star$) distribution of galaxies along the wavelength,
{ and in Fig.~\ref{Fig:px2} we give the photo-$z$ probability $P(\chi^2)$ (see Eq.~1) as a function of redshift for an arbitrary galaxy in the 2nd, 3rd and 4th subsamples in both cases of using $ugriz$ and $F_{\rm _{UV}}N_{\rm _{UV}}ugriz$ photometry, and in each panel of Fig.~\ref{Fig:px2} the magnitudes and magnitude-errors of the corresponding galaxy are also shown.}
Note that all of these magnitudes { are} corrected for foreground extinction by using the \citet{sch98} reddening maps, the extinction law of \citet{sch98} for SDSS { $ugriz$} magnitudes and that of \citet[hereafter CCM]{car89} for GALEX $F_{\rm _{UV}}$ and $N_{\rm _{UV}}$ magnitudes.
By averaging the CCM's extinction law over GALEX $F_{\rm _{UV}}$ and $N_{\rm _{UV}}$ bandpasses, \citet{wyd05} obtained $A_{\rm F_{\rm UV}}/E(B-V) = 8.374$ and $A_{\rm N_{\rm _{UV}}}/E(B-V) = 8.741$ \citep{don07}.
Meanwhile, all of these magnitudes { are} {\sl k-}corrected by using SDSS spectroscopic redshift $z_{\rm spec}$ and the $K$-correction v$4\_1\_4$ code of \citet{bla07}, in which the distance moduli are obtained by assuming the cosmological parameters $(\Omega_{\rm M}, \Omega_\Lambda)$= (0.3, 0.7).
\begin{enumerate}
\item {From Fig.~\ref{Fig:colrange} we see that} the galaxies in the 1st and 2nd subsamples are redder ($g^\star-r^\star \ga 0.70$) and bluer ($g^\star-r^\star \la 0.70$) { ones}, respectively. The main reason that the photo-$z$s of galaxies in the 2nd subsample are erroneously identified by using optical photometry is that { bluer galaxies are} often accompanied by star formation \citep{ric05}, { leading to} a probability function { $P(\chi^2)$} with significant secondary peak in the photo-$z$ calculation (see the left panel of Fig.~\ref{Fig:px2}).
    The secondary peak ({ in the regime} $1.8 \la z \la 2.8$) is { explained by that} the 4000\,$\rm \AA$ break, which is the most commonly used spectral feature for optical photo-$z$ determination, is greatly reduced { [}making it more difficult to use as a redshift indicator \citep[and references therein]{nie09}{ ]}, and { by that} in the $1.8 \la z \la 2.8$ region the 4000\,$\rm \AA$ break goes beyond the wavelength coverage of $ugriz$ filters { [}making it impossible to use as a redshift indicator{ ]}.
    When UV data is combined, the Lyman break, which is exhibited by all galaxies, would emerge in the UV ranges if $1.8 \la z \la 2.8$, this { would} largely { decrease} the probability of the solution in the high redshift domain.
\item the 3rd spectroscopic subsample (including 30 galaxies) is constituted of 9 galaxies with the full set of photometry and 21 galaxies with $N_{\rm _{UV}}ugriz$ detections (one has wrong photometry). 8 of the 9 galaxies with $F_{\rm _{UV}}N_{\rm _{UV}}ugriz$ detections and 15 of the 21 galaxies with $N_{\rm _{UV}}ugriz$ detections have fainter $u$-light ($\ga 20.5$\,mag) and larger $u$-magnitude errors (see Fig.~\ref{Fig:maglam} { and the middle panel of Fig.~\ref{Fig:px2}}), { and} therefore { have} bluer $F_{\rm _{UV}}^\star-u^\star$ and $N_{\rm _{UV}}^\star-u^\star$ colours and redder $u^\star-optical^\star$ colours (such as $u^\star-g^\star$ colour, see Fig.~\ref{Fig:colrange}).
    The larger $u$-magnitude error leads { to a smaller} contribution of the $u$-band photometry to the $\chi^2$ { calculation} (see Eq.~1), the probability function has two peaks for degeneracy of the fit parameters if only optical photometry { is used}.
    Even if the UV photometry is included{, the high redshift resolution} (at $z>2.3$, see Fig.~\ref{Fig:zcom-psb-UVv} { and the middle panel of Fig.~\ref{Fig:px2}) could not been excluded by the \textit{Hyperz} code. This is explained by that} the Lyman break would move to wavelength longer than 3000\,$\rm \AA$ [i.e., { emerge} in the $u$ band] { and the 4000\,$\rm \AA$\,break would go beyond the wavelength coverage of $ugriz$ filters}. Note that the
    confidence levels are very low in the determinations of { both} $z^{\rm V}_{\rm phot}$ and $z^{\rm UVV}_{\rm phot}$ for most of galaxies in this subsample (refer to Fig.~\ref{Fig:zcom-psb-UVv}).
\item the 4th subsample (9 galaxies) is constituted of 2 galaxies with the full set of photometry and 7 galaxies with $N_{\rm _{UV}}ugriz$ detections. The galaxies in this subsample have fainter $N_{\rm _{UV}}$ light { (see Fig.~\ref{Fig:maglam} and the right panel of Fig.~\ref{Fig:px2})}, therefore { have} redder $N_{\rm _{UV}}^\star-optical^\star$ colours (such as $N_{\rm _{UV}}^\star-u^\star$ colour, see Fig.~\ref{Fig:colrange}).
    Because the fainter $N_{\rm _{UV}}$ light also can be produced by high redshift galaxies, for which the Lyman break { would move} to the $N_{\rm _{UV}}$ passband{ \color{red} ,} the \textit{Hyperz} code is unable to decide between the high redshift and low redshift solutions. From the right panel of Fig.~\ref{Fig:px2} we see that the $P(\chi^2)$ of $z^{\rm V}_{\rm phot}$ is higher than that of $z^{\rm UVV}_{\rm phot}$.
\end{enumerate}

{ In summary, the addition of UV photometry would increase the number of non-catastrophic (i.e., $|\Delta z| = |z_{\rm spec} -z_{\rm phot}| < 1.0$) identifications, and decrease the mean deviation $\langle \Delta z \rangle$ ($\equiv \Sigma \Delta z/N$) and standard deviation $\sigma_z$ ($\equiv \sqrt{\Sigma(\Delta z-\langle \Delta z \rangle)^2/(N-1)}$) when all galaxies are included.
And, the 2nd spectroscopic subsample of galaxies, for which the photo-$z$ errors are largely decreased by the inclusion of UV data, mainly locates in the redshift $z_{\rm spec} <0.2$ and the colour $g^\star - r^\star \la 0.7$ regions, and has not significant difference in the UV and optical magnitudes in comparison with the 1st subsample.
Therefore, if the galaxy is a bluer one, we had better combine UV with optical photometry to obtain its photo-$z$ estimate.}
If the $N_{\rm _{UV}}$ light of a galaxy is too faint, { its photo-$z$ estimate would be identified erroneously when UV data is included}, and
if the $u$-light of a galaxy is too faint and the $u$-magnitude error is too large, { its photo-$z$ estimate still would be erroneously identified} even if UV photometry is included in the \textit{Hyperz} code.

\subsubsection{Dependence of photo-$z$ accuracy}
{ To investigate the performance of photo-$z$ estimate in removing catastrophic identifications, in Tables~\ref{Tab:perfor-z}-\ref{Tab:perfor-r} we give the number of non-catastrophic identifications $N$, mean deviation $\langle \Delta z \rangle$ and standard deviation $\sigma_z$ as a function of spectroscopic redshift $z_{\rm spec}$, $g-r$ colour and $r$-magnitude for Models A-D2. Each table includes the cases of using $ugriz$ and $F_{\rm _{UV}}N_{\rm _{UV}}ugriz$ photometry.
The results in Tables~\ref{Tab:perfor-z}-\ref{Tab:perfor-r} are plotted in Figs.~\ref{Fig:sigma-z}-\ref{Fig:sigma-r}, respectively.

Fig.~\ref{Fig:sigma-z} gives the evolutions of $N$, $\langle \Delta z \rangle$ and $\sigma_z$ with $z_{\rm spec}$ for Models A-D2.
\begin{itemize}
\item First, from the evolution of the mean deviation $\langle \Delta z \rangle$ with $z_{\rm spec}$ we see that if $ugriz$ photometry is used the mean photometric redshift $\langle z_{\rm phot} \rangle$ ($\equiv \Sigma z_{\rm phot}/N$) in the regions of $z_{\rm spec} \la 0.05$ and $0.25 \la z_{\rm spec} \la 0.45$ is greater ($\sim 0.05$ at the most), and that in the region of $0.05 \la z_{\rm spec} \la 0.25$ ($\sim 0.02$ at the most) is less than the mean spectroscopic redshift $\langle z_{\rm spec} \rangle$ ($\equiv \Sigma z_{\rm spec}/N$) for Models A, B, D and D2. For Models C and C2, $\langle z_{\rm phot} \rangle$ is greater than $\langle z_{\rm spec} \rangle$ at all redshift bins. The maximal $|\langle \Delta z \rangle|$ locates in the regions of $z_{\rm spec} \la 0.05$ and $0.25 \la z_{\rm spec} \la 0.45$ for all models. The addition of UV data almost does not vary the mean deviation except in the region $0.5 \la z_{\rm spec} \la 0.6$, in which the number of galaxies is smaller.

\item Secondly, from the correlation between $\sigma_z$ and $z_{\rm spec}$ we see that the standard deviation $\sigma_z$ in the $z_{\rm spec} \la 0.3$ region is less than that in the $z_{\rm spec} \ga 0.3$ region if only $ugriz$ data is used, the addition of UV data decreases $\sigma_z$ in the region of $z_{\rm spec} \ga 0.3$ for all models.

\item At last, from Fig.~\ref{Fig:sigma-z} we see that the addition of UV photometry significantly increases the number of non-catastrophic identifications in the low redshift domain for all models.
\end{itemize}

Fig.~\ref{Fig:sigma-gr} presents the evolutions of $N$, $\langle \Delta z \rangle$ and $\sigma_z$ as a function of $g-r$ colour, reveals the dependence of photo-$z$ accuracy on $g-r$ colour.
\begin{itemize}
\item
First, from the evolution of $\langle \Delta z \rangle$ with $g-r$ colour, we see that in the case of only using $ugriz$ photometry the mean deviation $\langle \Delta z \rangle$ is positive in the $0.8 \la g-r \la 1.4$ region, and in the $0.3 \la g-r \la 0.8$ region $\langle \Delta z \rangle$ is negative and $|\langle \Delta z \rangle|$ increases with decreasing $g-r$ colour for Models A, B, D and D2. For Models C and C2 the mean deviation $\langle \Delta z \rangle$ is less than zero at all $g-r$ color bins and $\langle \Delta z \rangle$ almost does not vary with $g-r$ colour. The inclusion of UV data can decrease $|\langle \Delta z \rangle|$ in the $0.3 \la g-r \la 0.8$ region for Models A, B, D and D2, while increase $|\langle \Delta z \rangle|$ at the redder $g-r$ colour for Models C and C2.

\item
Secondly, from the evolution of $\sigma_z$ with $g-r$ colour, we see that the standard deviation in the $0.8 \la g-r \la 1.2$ region is less than that in the other regions if $ugriz$ photometry is used, and the inclusion of UV
data would decrease the standard deviation in the regime $0.3 \la g-r \la 0.6$ (except for Model B) for all models.

\item
At last, from the correlation between $N$ and $g-r$ colour, we see that the addition of UV photometry increases the number of non-catastrophic identifications in the region $g-r \la 0.8$.

\end{itemize}

From  Fig.~\ref{Fig:sigma-r}, which gives the evolutions of $\langle \Delta z \rangle$, $\sigma_z$ and $N$ as a function of $r-$magnitude, we see that the mean deviation $\langle \Delta z \rangle$ and standard deviation $\sigma_z$ are independent of $r-$magnitude in both cases of using $ugriz$ and $F_{_{\rm UV}}N_{_{\rm UV}}ugriz$ photometry, the inclusion of UV photometry increases the number of non-catastrophic identifications in the range $17.0 \la r \la 18.5$, and increases $\sigma_z$ for all models.

In summary, if catastrophic identifications are removed, the inclusion of UV data raises the number of non-catastrophic identifications in the low redshift, $g-r \la 0.8$ and $17 \le r \le 18.5$ regions, decreases the mean deviation in the high redshift ($z_{\rm spec} \ga 0.5$) and $0.3 \la g-r \la 0.8$ regions, and also decreases the standard deviation in the $z_{\rm spec} \ga 0.3$ and $0.3\la g-r \la 0.6$ regions.}

\subsubsection{$F_{\rm _{UV}}$ or $N_{\rm _{UV}}$ photometry on photo-$z$ determinations}

If we only combine $F_{\rm _{UV}}$ or $N_{\rm _{UV}}$ with optical photometry, what results would happen? Using the set of $N_{\rm _{UV}}ugriz$ photometry, we obtain photo-$z$ estimates ($z{\rm_{phot}^{N_{\rm _{UV}}V}}$) of galaxies in the spectroscopic sample for Models A-D2, { then} compare them with the results obtained by using the set of $F_{\rm _{UV}}N_{\rm _{UV}}ugriz$ photometry (i.e., $z{\rm_{phot}^{UVV}}$).
For the sake of size, in the top panel of Fig.~\ref{Fig:lowz-dia-bc03-NUV} we only give a comparison between $z{\rm_{phot}^{UVV}}$ and $z{\rm_{phot}^{N_{\rm _{UV}}V}}$ for Model B. From it we see that for most of galaxies $z{\rm_{phot}^{N_{\rm _{UV}}V}}$ is lower than 1.0 and $z{\rm_{phot}^{N_{\rm _{UV}}V}} > z{\rm_{phot}^{UVV}}$.
That is to say, { only $N_{\rm _{UV}}$ is combined with $ugriz$ photometry}, the { derived} photo-$z$s  are reasonable for most of galaxies,
{ the number of non-catastrophic identifications $N$ is increased, but the mean deviation and standard deviations are raised in comparison with those in the case of using $F_{\rm _{UV}}N_{\rm _{UV}}ugriz$ photometry (due to $z{\rm_{phot}^{N_{\rm _{UV}}V}} >z{\rm_{phot}^{UVV}}$).}
The reason that the inclusion of $N_{\rm _{UV}}$ light decreases the probability of the solution { in the region of} $1.8 \la z \la 2.4$ is { that} the Lyman break would emerge in the $N_{\rm _{UV}}$ passband. The overestimation of photo-$z$ is { explained by that the main spectral feature still is the 4000\,$\rm \AA$\,break, which is unable to break the degeneracy among the fit parameters if $z \la 0.6$.}

Using the set of $F_{\rm _{UV}}ugriz$ photometry, we also obtain photo-$z$ estimates ($z{\rm_{phot}^{F_{\rm _{UV}}V}}$) for the spectroscopic galaxy sample{ . I}n the bottom panel of Fig.~\ref{Fig:lowz-dia-bc03-NUV} we give { a} comparison between $z{\rm_{phot}^{UVV}}$ and $z{\rm_{phot}^{F_{\rm _{UV}}V}}$ for Model B. From it we see that $z{\rm_{phot}^{F_{\rm _{UV}}V}}$ is close to $z{\rm_{spec}}$.
The agreement is caused by { the fact that} the Lyman break would emerge in the $F_{\rm _{UV}}$ passband if the redshift is { low}. Above conclusions are also true for the other Models.

{ In Tables~\ref{Tab:perfor-z}-\ref{Tab:perfor-r} we also give $N$, $\langle \Delta z \rangle$ and $\sigma_z$ of Model B as a function of redshift, $g-r$ and $r$-magnitude when using $N_{\rm _{UV}}ugriz$ or $F_{\rm _{UV}}ugriz$ photometry.
From them we see that the catastrophic objects mainly are those galaxies with $z_{\rm spec} \ga 0.3$, $g-r \ga 0.8$ and $r \ga 18$\,mag when $N_{\rm _{UV}}$ photometry is combined with optical photometry. Also, from them we see the catastrophic objects are those galaxies with $z_{\rm spec} \la 0.15$, $0.4 \la g-r \la 1.0$ and $16.5 \la r \la 18$\,mag if $F_{\rm _{UV}}$ photometry is combined with optical data.
That is to say, only including $N_{\rm _{UV}}$ or $F_{\rm UV}$ photometry would cause to erroneous photo-$z$ estimates for those faint 'redder' high-redshift galaxies and those 'bluer' low-redshift (intermediate $r$-magnitude) galaxies, respectively.
The number of catastrophic identifications in the case of using $F_{\rm UV}ugriz$ photometry is less than that of using $N_{\rm _{UV}}ugriz$ photometry.}

\subsection{Binary interactions on photo-$z$ determinations}
Binary interactions would significantly { change the UV spectra of populations at age} $t \ga 1$\,Gyr (see Fig.~\ref{Fig:isedcom-ath}). Using the HPL07-b, HPL07-s, Yunnan-b and Yunnan-s population synthesis models we get the $F_{\rm _{UV}},N_{\rm _{UV}},u,g,r,i$ and $z$ magnitudes of a solar-metallicity population with a mass of 1\,${\rm M}_\odot$. By comparison we find the differences in the $N_{\rm UV},u,g,r,i$ and $z$ magnitudes among { these} models { are slight}, while the difference in the $F_{\rm _{UV}}$ magnitude caused by binary interactions, i.e., that between HPL07-s and HPL07-b or that between Yunnan-s and Yunnan-b models, is significant at ${\rm log}(t/{\rm yr}) \ga 9.0$ (see Fig.~\ref{Fig:FUVcom}). Then whether binary interactions would affect photo-$z$ estimates?

Comparing the { photo-$z$ estimates} of galaxies in the spectroscopic sample between Models C and C2 (Fig.~\ref{Fig:lowz-dia-hpl}), and those between Models D and D2 (Fig.~\ref{Fig:lowz-dia-yn}), we find that binary interactions do not systematically affect photo-$z$ determinations.
One reason { for this} is that { binary interactions vary significantly the SEDs of populations (or Burst-type galaxies) only at late ages ($ t \ga 1$\,Gyr), leading that at redshift $z=0$ the difference in the SEDs, $F_{\rm _{UV}},N_{\rm _{UV}},u,g,r,i$ and $z$ magnitudes of non-Burst galaxies between Models C and C2 and that between Models D and D2 is slight at a given age and galaxy type. In Fig.~\ref{Fig:FUVcspcom} we give the $F_{\rm _{UV}}$ magnitude of Burst, E, S0 and Sa-Sd galaxies at redshift $z=0$ by using { the} Yunnan-s and Yunnan-b models.}
Even if { at redshift} $z=0.6$, binary interactions only make the ISEDs { of populations (or Burst-type galaxies) bluer in the regime $\lambda < 4000\,{\rm \AA}$}  { according to $z=(\lambda'-\lambda_0)/\lambda_0$ and $\lambda_0 = 2500\, \rm \AA$, beyond which the effect of binary interactions on the ISEDs is slight.
These bluer UV and near-optical ISEDs} can be { easily} offset by the decrements of visual extinction $A_{V}$ and age, absence of the Lyman break and the adoption of the different reddening laws during the fitting process.

{ In Figs.~\ref{Fig:sigma-z-models}-\ref{Fig:sigma-r-models} (also generated from Tables B1-B3 in turn), we present $N$, the absolute value of mean deviation $|\langle \Delta z \rangle|$ and $\sigma_z$ as a function of $z_{\rm spec}$, $g-r$ colour and $r-$magnitude for Models A-D2 in both cases of using $ugriz$ and $F_{\rm _{UV}}N_{\rm _{UV}}ugriz$ photometry, also the catastrophic identifications are removed.
From Figs.~\ref{Fig:sigma-z-models}-\ref{Fig:sigma-r-models} we see that binary interactions decrease and increase $|\langle \Delta z \rangle|$ in the regions of $0.3 \la g-r \la 0.8$ and $r \ga 17.8$\,mag in the case of using optical photometry, respectively (see the { comparisons} between Models D and D2 in Figs.~\ref{Fig:sigma-gr-models} and \ref{Fig:sigma-r-models}),
and decrease the standard deviation $\sigma_z$ at all $r$ magnitudes for both cases of using $ugriz$ and $F_{\rm _{UV}}N_{\rm _{UV}}ugriz$ data (see Fig.~\ref{Fig:sigma-r-models}).

In Fig.~\ref{Fig:dn-bi} we present the percent difference in the number of non-catastrophic identifications between Models C and C2 and that between Models D and D2 (${\rm d}N \equiv (N_{x2} - N_x)/N_{x2}$, $x$=C or D) as a function of $z_{\rm spec}$, $g-r$ colour and $r-$magnitude.
From it we see that $N_{\rm D2}$ is less than $N_{\rm D}$ at $0.2 \la z_{\rm spec} \la 0.3$ in both cases of using $ugriz$ ($\sim$10\% at the most) and $F_{\rm _{UV}}N_{\rm _{UV}}ugriz$ ($\sim$2\% at the most) photometry, greater than $N_{\rm D}$ in the region $0.3 \la g-r \la 0.8$ when only using optical data ($\sim 23$\% at the most), and almost equal to $N_{\rm D}$ in the whole $g-r$ region (except $g-r \ga 1.2$\,mag, in which $N$ is smaller) when using the full set of photometry.

In brief, the inclusion of binary interactions mainly affects the photo-$z$ determinations in the case of only using optical photometry, in an exact word, it increases the number of non-catastrophic identifications and decreases the deviations in the region of $0.3 \le g-r \le 0.8$\,mag if only optical photometry is used.}

\subsection{EPS models on photo-$z$ determinations}
{ First, from the evolutions of the number of non-catastrophic identifications $N$ with $z_{\rm phot}$, $g-r$ colour and $r-$magnitude in the top panels of Figs.~\ref{Fig:sigma-z-models}-\ref{Fig:sigma-r-models}, we see that the difference in $N$ among models A-D2 is smaller if $F_{\rm _{UV}}N_{\rm _{UV}}ugriz$ data is used, and see that $N$ decreases from Model B, Model A, Models D/D2 to C/C2 (i.e., $N_{\rm B} > N_{\rm A} > N_{\rm D/D2} > N_{\rm C/C2}$) in the regions of $0.05 \la z_{\rm spec} \la 0.15$, $0.3 \la g-r \la 0.8$ and $17 \la r \la 18.2$\,mag when only optical data is used.

Secondly, from the middle and bottom panels of Figs.~\ref{Fig:sigma-z-models}, we see that
(i) the absolute value of mean deviation $|\langle \Delta z \rangle|$ and standard deviation $\sigma_z$ of Models C/C2 are greater than those of the other models in the low redshift domain for both cases of using $ugriz$ and $F_{_{\rm UV}}N_{_{\rm UV}} ugriz$ photometry;
(ii) among Models A, B, D and D2, the difference of $|\langle \Delta z \rangle|$ in the regions of $z_{\rm spec} \la 0.05$ and $z_{\rm spec} \ga 0.5$ and the difference of $\sigma_z$ in the region of $z_{\rm spec} \ga 0.5$ are larger (except $\sigma_z$ in the case of using $F_{\rm _{UV}}N_{\rm _{UV}}ugriz$ data),
furthermore, in the $0.1 \la z_{\rm spec} \la 0.5$ region the $|\langle \Delta z \rangle|$ and $\sigma_z$ of Model B are greater than those of Models A, D and D2 for both cases of using $ugriz$ and $F_{\rm _{UV}}N_{\rm UV}ugriz$ photometry (except $\sigma_z$ by using optical data in the $0.3 \la z_{\rm spec} \la 0.4$ region);
(iii) the evolutions of $|\langle \Delta z \rangle|$ and $\sigma_z$ with redshift of Models D and D2 are similar to those of Model A.

Moreover, from the middle and bottom panels of Figs.~\ref{Fig:sigma-gr-models}, we see that in the $g-r \la 0.7$ region the $|\langle \Delta z \rangle|$ and $\sigma_z$ of these models (except $\sigma_z$ in the case of using $F_{\rm _{UV}}N_{\rm _{UV}}ugriz$ data) are significantly different, and that the $|\langle \Delta z \rangle|$ and $\sigma_z$ of Model B are significantly lower than those of the other models for both cases of using $ugriz$ and $F_{_{\rm UV}}N_{_{\rm UV}} ugriz$ photometry.

At last, from the middle and bottom panels of Figs.~\ref{Fig:sigma-r-models}, we see that the mean and standard deviations of Models C and C2 are greater than those of the other models in both cases of using $ugriz$ and $F_{_{\rm UV}}N_{_{\rm UV}} ugriz$ photometry, and in the intermediate $r$-magnitude region the $|\langle \Delta z \rangle|$ of Models A, D and D2 are lower in both cases of using $ugriz$ and $F_{_{\rm UV}}N_{_{\rm UV}} ugriz$ photometry.

From above comparisons, we see that in the case of using $F_{_{\rm UV}}N_{_{\rm UV}} ugriz$ photometry the number of non-catastrophic identifications is similar for all models, and in the case of using $ugriz$ photometry the number of non-catastrophic identifications of Models B is greater than that of the other models in the low redshift, bluer $g-r$ and faint $r-$magnitude regions. In both cases of using $ugriz$ and $F_{_{\rm UV}}N_{_{\rm UV}} ugriz$ photometry, the mean and standard deviations of Model B are lower at $g-r \la 0.7$, and $|\langle \Delta z \rangle|$ of Models A, D and D2 are relatively lower at almost all redshift bins and at intermediate $r$ magnitudes.}

\section{UV photometry and Binary interactions on galaxy morphology}
\begin{table}
\caption{Morphological index $T$ of \citet{fuk07} and Ours.}
\centering{
\begin{tabular}{lcc}
\hline
galaxy Type  & $T$(Fuk) & $T$(Ours) \\
\hline
Unclassified &   -1     &     ...   \\
 Burst       &  ...     &     1     \\
 E           &   0      &     2     \\
 S0          &   1      &     3     \\
 Sa          &   2      &     4     \\
 Sb          &   3      &     5     \\
 Sc          &   4      &     6     \\
 Sd          &   5      &     7     \\
 Irr         &   6      &     8     \\
 CWW-E       &  ...     &     9     \\
 CWW-Sbc     &  ...     &     10    \\
 CWW-Scd     &  ...     &     11    \\
 CWW-Irr     &  ...     &     12    \\
\hline
\end{tabular}
}
\label{Tab:def-typ}
\end{table}

\begin{table}
\caption{ {\sl Reliability} and {\sl completeness} for early- and late-type selection by the \textit{Hyperz} code (including Models A-D2).
The upper and lower parts correspond to using $ugriz$ and  $F_{\rm UV}N_{\rm _{UV}}ugriz$ photometry, respectively. This table is based on the morphology sample of 1,502 galaxies.}
\centering{
\begin{tabular}{l cc cc}
\hline
Model & \multicolumn{2}{c}{early type} & \multicolumn{2}{c}{late type} \\
       & {\sl reliability} & {\sl completeness} & {\sl reliability} & {\sl completeness} \\
\hline
\multicolumn{5}{c}{$ugriz$}\\
A    &   0.74   &   0.60   &   0.62   &   0.54\\
B    &   0.69   &   0.60   &   0.58   &   0.46\\
C    &   0.83   &   0.39   &   0.54   &   0.73\\
C2   &   0.85   &   0.40   &   0.54   &   0.74\\
D    &   0.72   &   0.58   &   0.60   &   0.53\\
D2   &   0.74   &   0.63   &   0.65   &   0.52\\

\multicolumn{5}{c}{$F_{\rm _{UV}}N_{\rm _{UV}}ugriz$}\\
A    &   0.70   &   0.58   &   0.57   &   0.49\\
B    &   0.70   &   0.58   &   0.57   &   0.49\\
C    &   0.86   &   0.43   &   0.57   &   0.74\\
C2   &   0.87   &   0.57   &   0.66   &   0.72\\
D    &   0.71   &   0.66   &   0.65   &   0.47\\
D2   &   0.71   &   0.67   &   0.66   &   0.45\\

\hline
\end{tabular}
}
\label{Tab:perf-t}
\end{table}

\begin{table}
\caption{ {\sl Reliability} and {\sl completeness} for
selecting early- and late-type galaxies by the three morphology
selection parameters: the concentration index $C$, profile
likelihood and $u-r$ colour.}
\centering{
\begin{tabular}{l cc}
\hline
Selection Rule & {\sl reliability}  & {\sl completeness} \\
\hline
\multicolumn{3}{c}{early type} \\
$C>2.6$                           &   0.83   &   0.74\\
$P_{\rm deV}>P_{\rm exp}$         &   0.76   &   0.77\\
$u-r\ge2.22$                      &   0.71   &   0.75\\

\multicolumn{3}{c}{         } \\
\multicolumn{3}{c}{late type} \\
$C<2.6$                           &   0.82   &   0.62\\
$P_{\rm deV}<P_{\rm exp}$         &   0.85   &   0.49\\
$u-r\le2.22$                      &   0.77   &   0.40\\

\hline
\end{tabular}
}
\label{Tab:perf-t2}
\end{table}

\begin{table}
\caption{ Similar to Table.~\ref{Tab:perf-t2}, but for
the four new constructed colour selection criteria.
}
\centering{
\begin{tabular}{l cc}
\hline
Selection Rule & {\sl reliability}  & {\sl completeness} \\
\hline
\multicolumn{3}{c}{early type} \\
$N_{\rm _{UV}}-u \ge 1.94$           &   0.89   &   0.65\\
$F_{\rm _{UV}}-u \ge 3.16$           &   0.88   &   0.37\\
$5.27-1.30(u-r) > (N_{\rm _{UV}}-u)$ &   0.88   &   0.65\\
$5.77-1.47(u-r) > (F_{\rm _{UV}}-u)$ &   0.89   &   0.64\\

\multicolumn{3}{c}{         } \\
\multicolumn{3}{c}{late type} \\
$N_{\rm _{UV}}-u < 1.94$             &   0.73   &  0.73\\
$F_{\rm _{UV}}-u < 3.16$             &   0.54   &  0.77\\
$5.27-1.30(u-r) < (N_{\rm _{UV}}-u)$ &   0.73   &  0.72\\
$5.77-1.47(u-r) < (F_{\rm _{UV}}-u)$ &   0.73   &  0.73\\

\hline
\end{tabular}
}
\label{Tab:perf-t-new}
\end{table}

\begin{figure}
\centering{
\includegraphics[bb=122 124 512 637,height= 5.5cm,width= 5.5cm,clip,angle=-90]{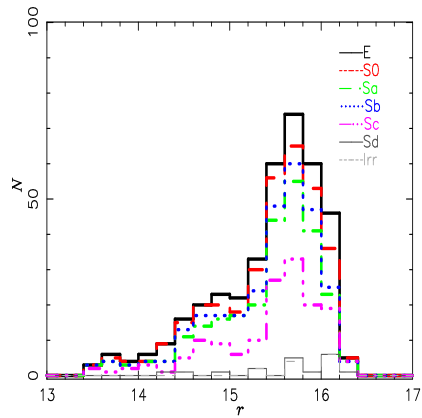}
}
\caption{ Histogram of $r$-magnitude distribution for
E (thick black solid line), S0 (thick red dashed), Sa (thick green
dot-dashed), Sb (thick blue dotted), Sc (thick magenta
dash-dot-dot-dot), Sd (thin grey solid) and Irr (thin grey dashed)
morphological subsample (see the text in Section 3). Note that the
half-integer classes are designated as the corresponding integer
classes in this figure.}
\label{Fig:his-mag}
\end{figure}

\begin{figure}
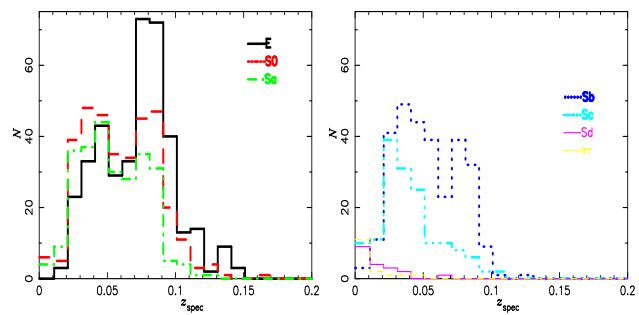

\centering{
\includegraphics[bb=122 124 520 637,height= 4.1cm,width= 4.2cm,clip,angle=-90]{his-z1.ps}
\includegraphics[bb=122 124 520 637,height= 4.1cm,width= 4.2cm,clip,angle=-90]{his-z2.ps}
}
\caption{ Similar to Fig.~\ref{Fig:his-mag}, but for
spectroscopic redshift $z_{\rm spec}$. Left panel is for E, S0 and
Sa types, right panel for Sb, Sc, Sd and Irr types. The lines have
the same meaning as in Fig.~\ref{Fig:his-mag}.}
\label{Fig:his-z}
\end{figure}

\begin{figure}
\centering{
\includegraphics[bb=136 134 513 625,height= 5.5cm,width= 5.5cm,clip,angle=-90]{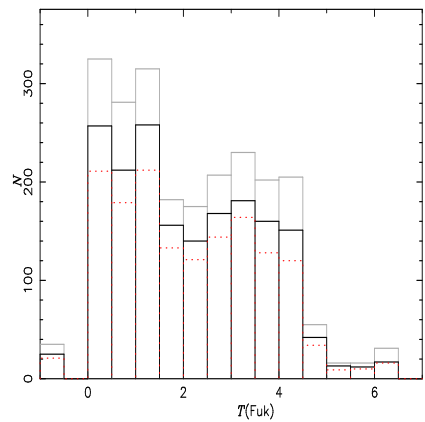}
}
\caption{ Histogram of morphology-index $T$(Fuk)
distribution for morphological galaxy sample. Grey solid line is
for the original catalogue of \citet{fuk07}, red dotted line is
for the sample of galaxies matched with the SDSS DR7 and GALEX DR4, also shown are the sample of galaxies only matched with the SDSS DR7 (black solid line). The corresponding galaxy types of $T$(Fuk) are given in Table ~\ref{Tab:def-typ}.}
\label{Fig:his-mor}
\end{figure}

\begin{figure}
\centering{
\includegraphics[bb=40 70 569 400,height= 7.5cm,width=11.5cm,clip,angle=-90]{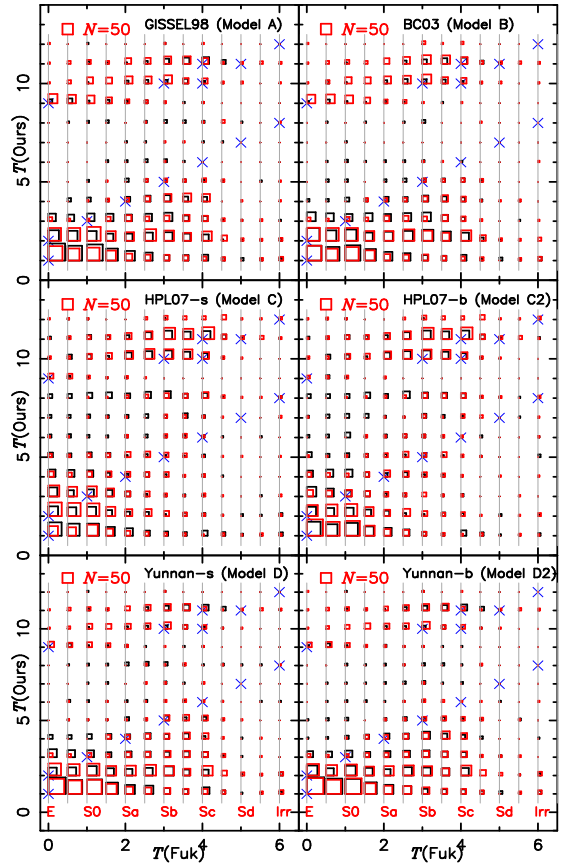}
}
\caption{Comparisons in the morphological index of galaxies between ours and those of Fukugita et al. for Models A-D2.
{ Black and red squares represent using $ugriz$ and
$F_{\rm _{UV}}N_{\rm _{UV}}ugriz$ photometry, respectively, and
their area represents the number of galaxies on the grid. The area
of square located in the upper left corner of each panel
denotes the number of galaxies is 50.}
Crosses show the correlation between $T$(Ours) and $T$(Fuk).}
\label{Fig:shape-all}
\end{figure}

\begin{figure}
\centering{
\includegraphics[bb=40 70 569 400,height= 7.5cm,width=11.5cm,clip,angle=-90]{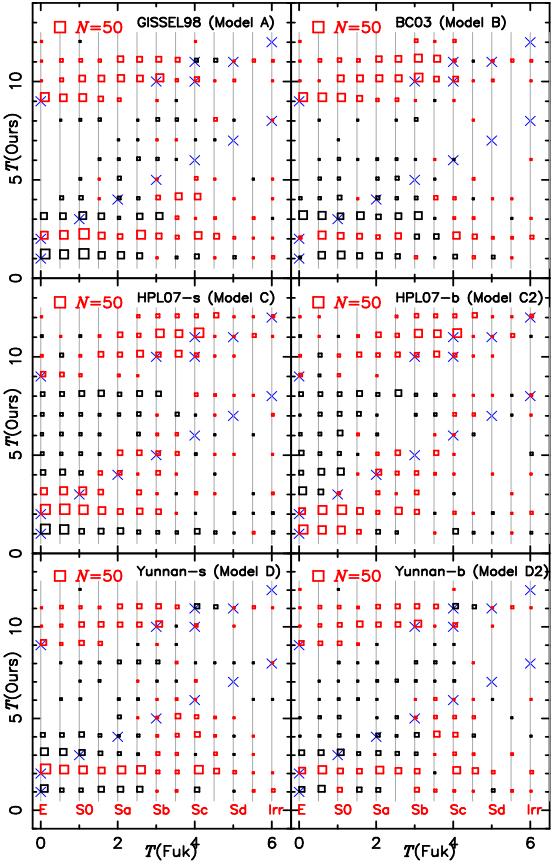}
}
\caption{ The difference in the number of galaxies between in the case of using $ugriz$ [$N(ugriz)$] and in the case of using $F_{_{\rm UV}}N_{_{\rm UV}}ugriz$ photometry [$N(F_{_{\rm UV}}N_{_{\rm UV}}ugriz)$] on the plane of $T$(Ours).vs.$T$(Fuk) for Models A-D2. Black and red symbols represent the difference d$N$ ($\equiv N(ugriz) - N(F_{_{\rm UV}} N_{_{\rm UV}}ugriz)$) is positive and negative, respectively. Also, The size of squares scales with the number of galaxies, and the area of square located in the upper left corner of each panel also denotes the number is 50. }

\label{Fig:dshape-all}
\end{figure}

\begin{figure}
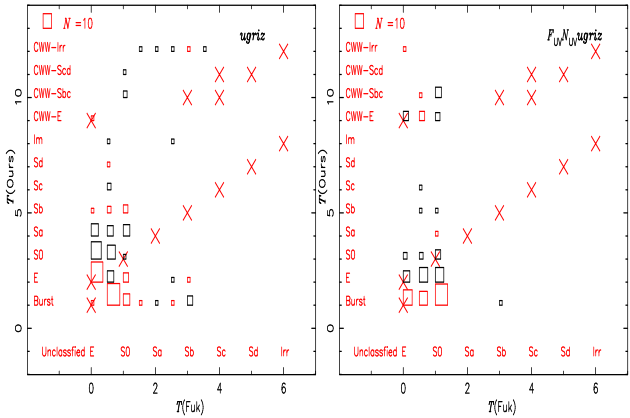

\centering{
\includegraphics[bb=136 134 513 625,height= 4.15cm,width= 5.5cm,clip,angle=-90]{94shape-dbi2-V.ps}\includegraphics[bb=136 134 513 625,height= 4.15cm,width= 5.5cm,clip,angle=-90]{94shape-dbi2-UVV.ps}
}
\caption{Similar to Fig.~\ref{Fig:dshape-all}, but { for the difference in the number of galaxies} between Models D and
D2 on the plane of $T$(Ours).vs.$T$(Fuk) (i.e., between neglecting and including binary interactions).
{Left and right panels are for using $ugriz$ and
$F_{\rm _{UV}}N_{\rm _{UV}}ugriz$ photometry, respectively.}
{ Red and black squares denote that the number of
galaxies of Model D on the grid is greater and less than that of Model D2, respectively. The size of symbols has the same meaning as in Fig.~\ref{Fig:dshape-all}.}}
\label{Fig:shapedbi-ath}
\end{figure}

\begin{figure}
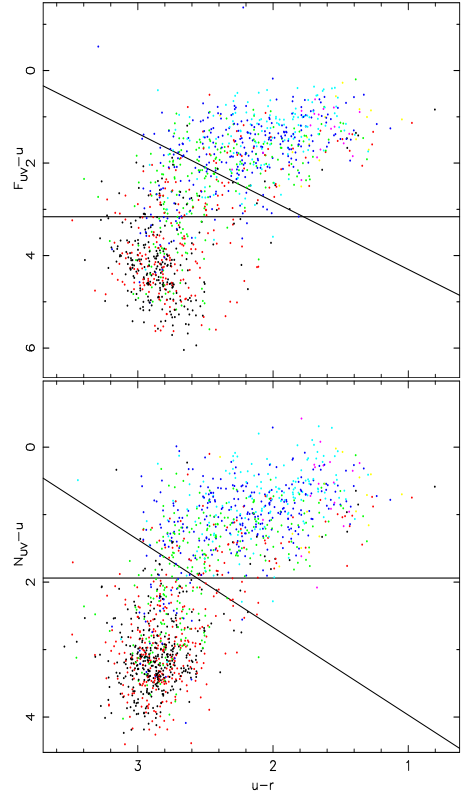

\centering{
\includegraphics[bb= 80 50 530 705,height= 6.0cm,width= 5.0cm,clip,angle=-90]{fuv-crit.ps}
\includegraphics[bb= 80 45 580 700,height= 6.0cm,width= 5.5cm,clip,angle=-90]{nuv-crit.ps}
}

\caption{ Distribution of galaxies in colour-colour diagrams (black, red, green, blue, cyaneous, yellow
and magenta symbols denote E, S0, Sa-Sd and Irr types,
respectively). Also shown are the lines corresponding to the four
criteria: $F_{\rm _{UV}}-u=3.16$ (horizontal line, top panel), $F_{\rm _{UV}}-u = 5.77-1.47(u-r)$ (oblique line, top panel), $N_{\rm _{UV}}-u=1.94$ (horizontal line, bottom panel) and $N_{\rm _{UV}}-u = 5.27-1.30(u-r)$ (oblique line, bottom panel).}
\label{Fig:mor-crit}
\end{figure}

\citet{bol00} claimed that the \textit{Hyperz} code allows to obtain photo-$z$ estimate and the best fit parameters across the whole space of galaxy{. B}ecause of the lack of spectral resolution, only { the rough spectral types} could be retrieved from broad-band photometry, { in an exact word,} early types could be reliably and easily identified while late types are usually misidentified because of the degeneracy between spectral type, age of the stellar population and visual extinction.

Using the { \textit{Hyperz} code and the} morphological sample of 1,502 galaxies, which are from the catalogue of \citet{fuk07} and { have} matched with the SDSS DR7 and GALEX DR4 (see the text in { Section} 3 for details), we { recover} their morphological types by using $ugriz$ and $F_{\rm _{UV}}N_{\rm _{UV}}ugriz$ photometry for Models A-D2, respectively. { The corresponding galaxy types of morphological index of \citet{fuk07}, $T$(Fuk), and ours, $T$(Ours), are presented in Table~\ref{Tab:def-typ}. Note that half-integer classes are allowed for $T$(Fuk).}

{ In Figs.~\ref{Fig:his-mag}-\ref{Fig:his-z} we give the histograms of $r-$magnitude and spectroscopic redshift $z_{\rm spec}$ distributions for E, S0, Sa-Sd and Irr subsamples, respectively, and in Fig.~\ref{Fig:his-mor} give the histogram of $T$(Fuk) distribution for the morphological sample. From them we see} that the galaxies in the morphological sample are relatively bright ($r < 16.${\,mag, see Fig.~\ref{Fig:his-mag}}) and their redshifts are less than 0.2 { (see Fig.~\ref{Fig:his-z})}, and that most of them belong to early types, Sd and Irr types are scarce { (see Fig.~\ref{Fig:his-mor})}.

\subsection{UV photometry on morphological types}
In Fig.~\ref{Fig:shape-all} we compare the retrieved morphological index $T{\rm (Ours)}$ with those of \citet{fuk07}, $T$(Fuk), for Models A-D2, note that those galaxies with $|z_{\rm phot}-z_{\rm spec}| > 1$ are excluded.

From the comparisons in Fig.~\ref{Fig:shape-all} we indeed see that in the case of only using optical photometry early types ($T {\rm (Fuk)} < 2.0$) can be easily retrieved and late types ($T {\rm (Fuk)} \ge 2.0$) are often misidentified as early types for all models except for Models C-C2. For Models C-C2 both early and late types can be reliably identified. And, if $F_{\rm _{UV}}N_{\rm _{UV}}ugriz$ photometry is used the situation { has not changed significantly.

Moreover, from Fig.~\ref{Fig:dshape-all}, which gives the difference in the number of galaxies between in the cases of using $ugriz$ and $F_{\rm _{UV}}N_{\rm _{UV}}ugriz$ photometry on the plane of $T$(Ours).vs.$T$(Fuk), we see that the probability that  early types ($T$(Fuk) $\le 2.0$) are classified as Burst and E (and CWW-E for Models A-B) types is decreased and increased, respectively, and that the probability that late types ($T$(Fuk)$> 2.0$) are classified as CWW-Sbc and CWW-Scd types is increased.}

{ To investigate the effect of UV photometry on the determinations of morphological types and the performance of the retrieved morphology by the \textit{Hyperz} code, we define $\langle \Delta T_i \rangle \equiv \Sigma _{k=1}^{N_i} [T({\rm Ours})-T({\rm Fuk})]/N_i$ and
$\sigma_{T_i} \equiv \sqrt{\Sigma_{k=1}^{N_i} [T({\rm Ours})-T({\rm Fuk})-\langle \Delta T_i \rangle]^2 /(N_i-1)}$,
in which $N_i$ is the number of non-catastrophic objects (i.e., $|z_{\rm spec}-z_{\rm phot}| < 1.$) classified as $i$-type by \citet[][i.e., $T$(Fuk)=$i$]{fuk07} and $i$ is allowed to be half-integer.
In Figs.~\ref{aFig:z}-\ref{aFig:r} we present the evolutions of $|\langle \Delta T_i \rangle|$, $\sigma_{T_i}$ and $N_i$ as a function of spectroscopic redshift $z_{\rm spec}$, $g-r$ colour and $r$-magnitude for Models A-D2.
}

{
The following comparisons in $|\langle \Delta T_i \rangle|$ and $\sigma_{T_i}$ are based on those bins with $N_i \ge 10$ because the small value in $N_i$ would lead to a larger error in $|\langle \Delta T_i \rangle|$ and $\sigma_{T_i}$. In each panel of Figs.~\ref{aFig:z}-\ref{aFig:r} the two vertical lines indicate the place of $N_i=10$, therefore we compare $|\langle \Delta T_i \rangle|$ and $\sigma_{T_i}$ in the region enclosed by two horizontal lines (shaded region of Figs.C1-C3).

\begin{itemize}
\item First, from Fig.~\ref{aFig:z}, we see that the inclusion of UV light increases $N_i$ of all models for $T$(Fuk)$\ge 1.0$ [the increment increases with $T$(Fuk)], decreases $|\langle \Delta T_i \rangle|$ and $\sigma_{T_i}$ of Models C and C2 for $T$(Fuk)$=0.0-0.5$ at all redshift bins and for $1.0 \le T$(Fuk)$\le 3.0$ (except $=2.5$) at $z_{\rm spec} \la 0.04$, and also decreases $|\langle \Delta T_i \rangle|$ and $\sigma_{T_i}$ of Models A and B for $T$(Fuk)=2.0-2.5 at all redshift bins.
    Moreover, we also see that all models have the similar trends of $|\langle \Delta T_i \rangle|$ and $\sigma_{T_i}$ with redshift for a given $T$(Fuk) when the full set of photometry is used: both $|\langle \Delta T_i \rangle|$ and $\sigma_{T_i}$ are independent of $z_{\rm spec}$ for all galaxy types, and $|\langle \Delta T_i\rangle|$ increases with $T$(Fuk) [except Models C and C2, for which $|\langle \Delta T_i\rangle|$ is independent of $T$(Fuk)].
\item Secondly, from Fig.~\ref{aFig:gr}, we see that the inclusion of UV data increases $N_i$ of all models for $2.5 \le T$(Fuk)$\le 4.0$ in the blue $g-r$ region [correspondingly, the left vertical line moves bluewards significantly, and $N_i$ increases with $T$(Fuk)], decreases $|\langle \Delta T_i \rangle|$ and $\sigma_{T_i}$ of Models C and C2 at $T$(Fuk)$\le 2.5$ and decreases $|\langle \Delta T_i \rangle|$ and $\sigma_{T_i}$ of all models at $T$(Fuk)$=3.0$ in the region of $g-r \la 0.8$\,mag.
    Moreover, if UV data is used, $|\langle \Delta T_i \rangle|$ correlates with $g-r$ colour and $T$(Fuk): $|\langle \Delta T_i \rangle|$ increases with redder $g-r$ ($\ga 0.8$) colour for $T$(Fuk)$\ga$2.0 (i.e, late types) and increases with $T$(Fuk) for all models except Models C and C2.
\item At last, from Fig.~\ref{aFig:r}, we see that the inclusion of UV data increases $N_i$ of all models for late types in the region of $r \ga 14\,$mag, decreases $|\langle \Delta T_i \rangle|$ and $\sigma_{T_i}$ of Models C and C2 for $T$(Fuk)$\le1.0$ at all $r$ magnitudes, and decreases $\sigma_{T_i}$ and increases $|\langle \Delta T_i \rangle|$ of Models A and B at $2.0 \le T$(Fuk)$\le 3.5$ at almost all $r$ magnitudes.
    The standard deviation $\sigma_{T_i}$ is relatively large for fainter ($r\ga16\,$mag) galaxies in the case of using $ugriz$ data, when UV photometry is included $\sigma_{T_i}$ almost does not vary with increasing $r$-magnitude except in the region of $r\ga 16$\,mag.
\end{itemize}
}

\subsection{Binary interactions on morphological types}
To obtain the influence of binary interactions on the retrieved morphological index, in Fig.~\ref{Fig:shapedbi-ath} we give a difference in the number of non-catastrophic objects between Models D and D2 on { the plane of $T$(Ours) and $T$(Fuk) for} $ugriz$ and $F_{\rm _{UV}}N_{\rm _{UV}}ugriz$ photometry.

From Figs.~\ref{Fig:shape-all} and \ref{Fig:shapedbi-ath}, we see that the inclusion of binary interactions mainly affects the determinations of E and S0 types { in both cases of using optical and the full set of photometry, and the difference in the case of using optical photometry is more significant than that in the case of using $F_{\rm _{UV}}N_{\rm _{UV}}ugriz$ photometry. This is explained by that binary interactions make the UV flux of Burst-type galaxies higher only at log($t$/yr)$\ga 9$, and these bluer UV spectra can be offset by the variation of other parameters (such as the decrements of age and visual extinction, etc) during the fitting process (also see the discussion in Section 4.2).}

\subsection{Comparison with other morphology selection criteria}
{ In this section we will give a comparison in the performance of selecting early- (E, S0 and Sa) and late-type (Sb, Sc, Sd and Irr) galaxies between by using the \textit{Hyperz} code (this work) and by using the three morphology-sensitive parameters (the concentration index $C$, the profile likelihood and $u-r$ colour).}
The concentration index $C$ is defined as the ratio of the radii containing 90\% and 50\% of the \textit{Petrosian} $r$ galaxy light, i.e., $C\equiv r_{p90}/r_{p50}$.
The profile likelihood is defined as the difference between the deVaucouleur's ($P_{\rm deV}$) and exponential ($P_{\rm exp}$) profile likelihoods in the $r$ band.

{ In Tables.~\ref{Tab:perf-t} and ~\ref{Tab:perf-t2} we give the {\sl reliability} and {\sl completeness} for selecting early- and late-type galaxies by using the \textit{Hyperz} code (including Models A-D2) and by using the three morphology selection criteria ($C=2.6$, $P_{\rm exp}-P_{\rm deV}=0$ and $u-r=2.22$).
The definitions of {\sl reliability} and {\sl completeness} are from \citet{str01}, i.e., the {\sl reliability} of classification is the fraction of galaxies from the selected subsample that are correctly classified, and the {\sl completeness} is the fraction of all galaxies of a given type from the original sample that are selected by the classification scheme.

From Tables.~\ref{Tab:perf-t} and ~\ref{Tab:perf-t2} we see that the {\sl reliability} for early-type selection by the \textit{Hyperz} code (including all Models) is similar to that by the three morphology selection criteria, while the {\sl completeness} is relatively low. And, the {\sl reliability} and {\sl completeness} for selecting late-type galaxies (except Models C and C2) by the \textit{Hyperz} code is less than those of the three morphology criteria.}

\subsection{New morphology selection criteria}
{ Plotting the galaxies in the morphological sample in colour-colour diagrams (Fig.~\ref{Fig:mor-crit}), we find that $F_{\rm _{UV}}-u$ and $N_{\rm _{UV}}-u$ can be used as new morphology indicators. In Table~\ref{Tab:perf-t-new} we present the {\sl reliability} and {\sl completeness} for selecting early- and late-type galaxies by the $F_{\rm _{UV}}-u = 3.16$ and $N_{\rm _{UV}}-u = 1.94$ separators.
We find that the $N_{\rm _{UV}}-u = 1.94$ colour separator has comparable {\sl reliability} and {\sl completeness} for both early- and late-type selection to the concentration index criterion ($C=2.6$), and has higher {\sl completeness} ($\sim$73\%) for late-type selection than the $u-r=2.22$ colour separator ($\sim$40\%).
Moreover, the $F_{\rm _{UV}}-u = 3.16$ colour separator has less {\sl completeness} for selecting early-type galaxies and less {\sl reliability} for selecting late-type galaxies than $C=2.6$, profile likelihood $P_{\rm deV} = P_{\rm exp}$ and $u-r=2.22$ separators, this is explained by that Sa-type galaxies show more scatter in $F_{\rm _{UV}}-u$ colour.

If the galaxies have not been detected in the $N_{\rm _{UV}}$ passband, we can construct the morphology selection criterion by combining their optical with $F_{\rm _{UV}}$ photometry, such as, we construct a morphology selection separator of $5.77-1.47(u-r)-(F_{\rm _{UV}}-u)=0$. By comparison we find that this separator has comparable {\sl reliability} and {\sl completeness} for both early- and late-type selection to $C=2.6$ and $N_{\rm _{UV}}-u=1.94$ selection criteria (see Table.~\ref{Tab:perf-t-new}) and has higher {\sl completeness} for selecting late types than $u-r=2.22$ separator.
Similarly, we also construct the $5.27-1.30(u-r)-(N_{\rm _{UV}}-u) =0$ separator, and find that this criterion has similar {\sl reliability} and {\sl completeness} for both early- and late-type selection to $N_{\rm UV}-u=1.94$ criterion, i.e., the performance of early- and late-type selection has not been improved even if the optical photometry is combined with $N_{\rm _{UV}}-u$ colour.
In Fig.~\ref{Fig:mor-crit} we present the lines, corresponding to the morphology selection separators, in the diagrams of $(F_{\rm UV}-u)$.vs.$(u-r)$ and $(N_{\rm _{UV}}-u)$.vs.$(u-r)$.}

\section{Summary and Conclusions}
Using the \textit{Hyperz} code of \citet{bol00} and a template spectral library which consists of 4 observed galaxy spectra from CWW and 8 spectral families built with the GISSEL98, BC03, HPL07 and Yunnan EPS models, respectively, we present photometric redshift estimates for a spectroscopic sample of galaxies which are selected randomly from the SDSS DR7 and GALEX DR4, and { present} morphological types for a morphological sample of bright galaxies which are also matched with the SDSS DR7 and GALEX DR4.

Based on the spectroscopic galaxy sample, we { find that the inclusion of $F_{\rm _{UV}}$ or $N_{\rm _{UV}}$ or both photometry can decrease the number of catastrophic identifications ($|z_{\rm spec}-z_{\rm phot}| > 1.0$), and the number of catastrophic identifications decreases from in the case of using $F_{\rm _{UV}}N_{\rm _{UV}}ugriz$ to $F_{\rm UV}ugriz$ and $N_{\rm UV}ugriz$ photometry.
In the case of using $F_{\rm _{UV}}N_{\rm _{UV}}ugriz$ photometry the catastrophic objects are those galaxies with { blue $g-r$ colour} and low spectroscopic redshift, in the case of using $N_{\rm _{UV}}ugriz$ photometry they are those faint 'redder' high-redshift galaxies, and in the case of using $F_{\rm _{UV}}ugriz$ photometry the catastrophic identifications are those 'bluer' low-redshift galaxies. If GALEX $F_{\rm _{UV}}$ photometry or both $F_{\rm _{UV}}$ and $N_{\rm _{UV}}$ photometry are combined with $ugriz$ data, the difference between photometric and spectroscopic redshifts is within 0.2 for the majority of non-catastrophic objects.
If} only $N_{\rm _{UV}}$ is combined, the derived photo-$z$s are systematically greater than their spectroscopic counterpart.

{ If catastrophic identifications are removed, the inclusion of both $F_{\rm UV}$ and $N_{\rm UV}$ photometry can raise the number of non-catastrophic identifications in the low redshift, $g-r \la 0.8$ and faint $r$-magnitude regions, decrease the mean deviation $|\langle \Delta z \rangle|$ in the $z_{\rm spec} \ga 0.5$ and $0.3 \la g-r \la 0.8$ regions (except Models C and C2), and decrease the standard deviation $\sigma_z$ in the $z_{\rm spec} \ga 0.3$ and $0.3 \la g-r \la 0.6$ regions.
Moreover, if a galaxy has fainter $u$ light, its photo-$z$ estimate still would be erroneously identified even if UV photometry is included, and if a galaxy has fainter $N_{\rm _{UV}}$ light, its photo-$z$ estimate would be erroneously identified when UV photometry is included.
Binary interactions mainly can raise the number of non-catastrophic identifications and decrease the mean and standard deviations in the 0.3$\le g-r \le 0.8$ region in the case of only using optical photometry.}

Based on the morphological galaxy sample, we { confirm that early types can be easily retrieved and late types are often misidentified as early types, and} find that the inclusion of UV photometry { decreases and increases the probability that early types are classified as Burst and E types, respectively, and increases the probability that late types are classified as CWW-Sbc and CWW-Scd types.
If catastrophic identifications are excluded, the inclusion of UV data raises the number of the retrieved late-type galaxies in all redshift, bluer $g-r$ and $r \ga 14$ regions, and the increment increases with $T$(Fuk).}
Binary interactions mainly affect the determination of E and S0 types.

{ By comparison we find the {\sl reliability} and {\sl completeness} for early- and late-type selection by the \textit{Hyperz} code are much lower than those by the three morphology criteria: concentration index $C=2.6$, profile likelihood $P_{\rm deV} = P_{\rm exp}$ and $u-r=2.22$.
Moreover, we find that $N_{\rm _{UV}}-u = 1.94$ can be used as a new morphology selection indicator, its {\sl reliability} and {\sl completeness} for selecting early and late types are approximately the same as those of $C=2.6$ criterion, and even its {\sl completeness} for selecting early types is greater than that of $u-r=2.22$ criterion.
The criterion of $F_{\rm _{UV}}-u = 3.16$ is not as good as $N_{\rm UV}-u = 1.94$, while the criterion constructed by the combination of $F_{\rm _{UV}}-u$ with $u-r$ colour, $5.77-1.47(u-r) = F_{\rm _{UV}}-u$, can reach to the similar {\sl reliability} and {\sl completeness} for early- and late-type selection to those of $N_{\rm _{UV}}-u=1.94$.
For the new constructed criteria $5.27-1.30(u-r) = N_{\rm _{UV}}-u$, the {\sl reliability} and {\sl completeness} for early- and late-type selection are not been improved in comparison with $N_{\rm _{UV}}-u = 1.94$ criterion.}

\section*{acknowledgments}
This work was funded by the Chinese Natural Science Foundation (Grant Nos 10773026, 10673029, 10821026 \& 2007CB15406) and by Yunnan Natural Science Foundation (Grant No 2007A113M). We are also grateful to the referee for suggestions that have improved the quality of this manuscript.

Funding for the SETS and SDSS-II has been provided by the Alfred P. Sloan Foundation, the Participating Institutions, the National Science Foundation, the U.S. Department of Energy, the National Aeronautics and Space Administration, the Japanese Monbukagakusho, the Max Planck Society, and the Higher Education Funding Council for England. The SDSS Web Site is http://www.sdss.org/.

The SDSS is managed by the Astrophysical Research Consortium for the Participating Institutions. The Participating Institutions are the American Museum of Natural History, Astrophysical Institute Potsdam, University of Basel, University of Cambridge, Case Western Reserve University, University of Chicago, Drexel University, Fermilab, the Institute for Advanced Study, the Japan Participation Group, Johns Hopkins University, the Joint Institute for Nuclear Astrophysics, the Kavli Institute for Particle Astrophysics and Cosmology, the Korean Scientist Group, the Chinese Academy of Sciences (LAMOST), Los Alamos National Laboratory, the Max-Planck-Institute for Astronomy (MPIA), the Max-Planck-Institute for Astrophysics (MPA), New Mexico State University, Ohio State University, University of Pittsburgh, University of Portsmouth, Princeton University, the United States Naval Observatory, and the University of Washington.

The galaxy Evolution Explorer (GALEX) is a NASA Small Explorer. The mission was developed in cooperation with the centre National d'Etudes Spatiales of France and the Korean Ministry of Science and Technology.

{}

\appendix
\section{The influence of input parameters on photo-$z$ determinations.}
In this section our standard is Model B. To investigate the effect of the input parameters on photo-$z$ estimates we proceed by changing each parameter, in turn, while the other parameters remain the same as in the standard model.
Model Ba uses the reddening law of \citet{sea79} instead of \citet{cal00}{ .} Model Bb uses the set of minimum
magnitudes [26, 26, 24, 24, 24, 24, 24] instead of [29.5, 29.5, 30, 30, 30, 30, 30]{ .} Models Bc and Bd use the minimum magnitude error of 0.001 and 0.10 instead of 0.05{ .} Models Be and Bf only use the 8 theoretical spectral families and CWW spectra, respectively{ .} Model Bg uses the set of cosmological parameters [1.0, 0.0, 50] instead of [0.3, 0.7, 70].

\begin{figure}
\centering{
\includegraphics[bb=79 86 572 597,height=6.50cm,width= 7.00cm,clip,angle=-90]{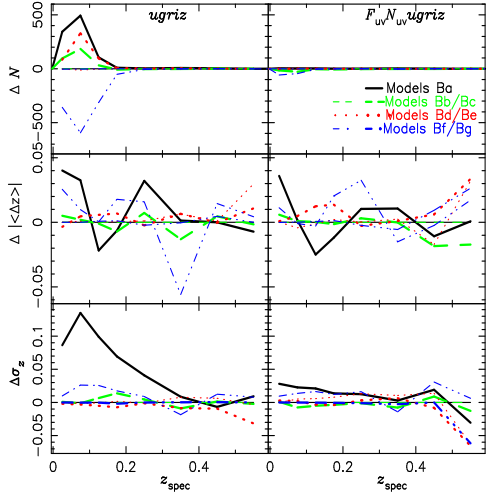}
}
\caption{ The differences in $N$ (top panels), $|\langle \Delta z \rangle|$ (middle panels) and $\sigma_z$ (bottom panels) between Model B and the other models Ba-Bg as a function of $z_{\rm spec}$ in removing catastrophic identifications. Left and right panels are for using $ugriz$ and $F_{\rm _{UV}}N_{\rm _{UV}}ugriz$ photometry, respectively.
}
\label{Fig:dsigma-inp-z}
\end{figure}

\begin{figure}
\centering{
\includegraphics[bb=79 86 572 597,height=6.50cm,width= 7.00cm,clip,angle=-90]{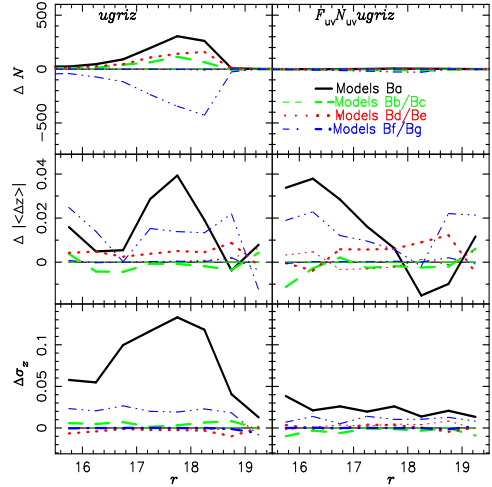}
}
\caption{ Similar to Fig.~\ref{Fig:dsigma-inp-z}, but a function of $r-$magnitude.}
\label{Fig:dsigma-inp-r}
\end{figure}

We obtain photo-$z$ estimates of galaxies in the spectroscopic sample for Models Ba-Bf { by using $ugriz$ and $F_{\rm _{UV}}N_{\rm _{UV}}ugriz$ photometry}.
{ In Fig.~\ref{Fig:dsigma-inp-z}-\ref{Fig:dsigma-inp-r} we present the differences in the number of non-catastrophic identifications ($\Delta N$), mean ($\Delta |\langle \Delta z \rangle|$) and standard deviations ($\Delta \sigma_z$) between Model B and the other models Ba-Bg as a function of redshift, $g-r$ colour and $r$-magnitude in removing catastrophic identifications, respectively. From them we see that}\\
{
$\bullet$ if the reddening law of \citet{sea79} [Model Ba] is used, the number of non-catastrophic identifications $N$ increases in the case of using optical photometry, the mean and standard deviations (except in the case of using $F_{\rm _{UV}}N_{\rm _{UV}}ugriz$ photometry) significantly differ from those of Model B;\\
$\bullet$ if the minimum magnitude error is set to 0.001 (Model Bc, one-fifty of the value in Model B) the number of non-catastrophic identifications $N$, the mean and standard deviations in the bluer $g-r$ region are greater than those of Model B in the case of only using optical photometry;\\
$\bullet$ if only theoretical spectral template (Model Be) or only the CWW set of empirical spectra (Model Bf) is used, the number of non-catastrophic identifications would be increased or decreased in the case of using optical photometry, respectively, the standard deviation is not changed significantly,
and the mean deviation of Model Bf is greater than that of Model Be; \\
$\bullet$ decreasing the set of minimum magnitude (Model Bb) or increasing the minimum magnitude error by a factor of 2 (0.1, Model Bd) or using the set of cosmological parameters of [1.0, 0., 70] would not vary significantly the results.\\
}


\begin{figure}
\centering{
\includegraphics[bb=79 86 572 597,height=6.50cm,width= 7.00cm,clip,angle=-90]{inp-dgr}
}
\caption{ Similar to Fig.~\ref{Fig:dsigma-inp-z}, but a function of $g-r$ colour.}
\label{Fig:dsigma-inp-gr}
\end{figure}

\section{Dependence of photo-$z$ accuracy on redshift, colour, magnitude and the filter set.}
In Tables B1-B3 we give the number of non-catastrophic
identifications $N$, standard deviation $\sigma_z$ and mean
deviation $\langle \Delta z \rangle$ as a function of
spectroscopic redshift $z_{\rm spec}$, $g-r$ colour and
$r$-magnitude in removing catastrophic identifications for Models
A-D2. Each table includes the cases of using $ugriz$, $F_{\rm
_{UV}}N_{\rm _{UV}}ugriz$,  $F_{\rm _{UV}}ugriz$ (only for Model
B) and $N_{\rm _{UV}}ugriz$ (only for Model B) photometry.

\begin{table*}
\tiny \caption{The number of non-catastrophic identifications $N$,
standard deviation ($\sigma_z$) and mean deviation ($\langle
\Delta z \rangle $) of galaxies in the spectroscopic sample as a
function of the redshift bin and the filter set for Models A-D2.
{ The first line gives the number of galaxies in the
corresponding redshift bin}. The following 1st and 2nd parts give
the results when { $ugriz$ and $F_{\rm _{UV}}N_{\rm _{UV}}ugriz$ photometry is used} for all models, { and the 3rd and 4th parts give the results when $N_{\rm _{UV}}ugriz$ and $F_{\rm _{UV}}ugriz$ photometry is used for Model B. All results are in the cases of removing catastrophic identifications.}}
\begin{tabular}{lcc ccc ccc}
\hline
          & $z=0.0-0.05$ & $z=0.05-0.10$ &$z=0.10-0.15$ &$z=0.15-0.20$ &$z=0.20-0.30$ &$z=0.30-0.40$ &$z=0.40-0.50$ &$z=0.50-0.60$ \\
        & $N,\sigma_z,\langle\Delta z\rangle$ & $N,\sigma_z,\langle\Delta z\rangle$ & $N,\sigma_z,\langle\Delta z\rangle$ & $N,\sigma_z,\langle\Delta z\rangle$ & $N,\sigma_z,\langle\Delta z\rangle$ & $N,\sigma_z,\langle\Delta z\rangle$ & $N,\sigma_z,\langle\Delta z\rangle$ & $N,\sigma_z,\langle\Delta z\rangle$\\
\hline
All &1114&2395&1796& 783& 288& 101&  47&   5\\

\multicolumn{9}{c}{ $ugriz$, non-cata.}\\
 A&
 722, 0.054,-0.035&1658, 0.041, 0.003&1573, 0.045, 0.017& 765, 0.045, 0.011&
 279, 0.056, 0.004&  87, 0.085,-0.047&  41, 0.082, 0.017&   5, 0.064,-0.032\\
 B&
 741, 0.054,-0.022&1881, 0.040, 0.012&1680, 0.045, 0.025& 771, 0.046, 0.014&
 279, 0.064,-0.007&  87, 0.077,-0.061&  41, 0.086,-0.001&   5, 0.066,-0.022\\
 C&
 673, 0.061,-0.060&1471, 0.058,-0.022&1407, 0.060,-0.022& 724, 0.055,-0.027&
 273, 0.062,-0.036&  86, 0.069,-0.058&  40, 0.086,-0.016&   5, 0.067,-0.019\\
 C2&
 666, 0.061,-0.060&1464, 0.059,-0.021&1405, 0.061,-0.020& 723, 0.056,-0.025&
 271, 0.061,-0.038&  80, 0.069,-0.065&  40, 0.086,-0.016&   4, 0.058,-0.047\\
 D&
 693, 0.058,-0.042&1567, 0.044, 0.001&1538, 0.049, 0.012& 756, 0.048, 0.008&
 278, 0.058, 0.002&  87, 0.085,-0.050&  41, 0.092, 0.015&   5, 0.104, 0.003\\
 D2&
 706, 0.055,-0.039&1609, 0.042, 0.003&1550, 0.046, 0.015& 738, 0.046, 0.011&
 250, 0.058,-0.005&  81, 0.082,-0.052&  35, 0.086, 0.016&   4, 0.051,-0.050\\

\multicolumn{9}{c}{ $F_{\rm _{UV}} N_{\rm _{UV}}ugriz$, non-cata.}\\
 A&
1109, 0.056,-0.027&2385, 0.049, 0.009&1790, 0.052, 0.021& 780, 0.049, 0.016&
 283, 0.059,-0.001&  94, 0.073,-0.040&  44, 0.050, 0.015&   5, 0.080, 0.019\\
 B&
1109, 0.055,-0.023&2384, 0.049, 0.016&1790, 0.055, 0.027& 781, 0.056, 0.014&
 283, 0.064,-0.008&  93, 0.081,-0.050&  44, 0.049, 0.018&   5, 0.081, 0.021\\
 C&
1111, 0.066,-0.043&2382, 0.064,-0.021&1791, 0.066,-0.027& 780, 0.064,-0.035&
 283, 0.069,-0.051&  92, 0.071,-0.070&  42, 0.066,-0.040&   5, 0.067,-0.053\\
 C2&
1110, 0.068,-0.045&2378, 0.065,-0.020&1791, 0.068,-0.025& 779, 0.065,-0.033&
 283, 0.070,-0.052&  90, 0.069,-0.069&  42, 0.066,-0.032&   4, 0.063,-0.070\\
 D&
1109, 0.061,-0.038&2383, 0.053, 0.001&1790, 0.053, 0.013& 780, 0.051, 0.014&
 283, 0.058,-0.003&  93, 0.079,-0.044&  44, 0.048, 0.020&   5, 0.103, 0.038\\
 D2&
1109, 0.061,-0.036&2381, 0.052, 0.003&1788, 0.052, 0.015& 777, 0.051, 0.012&
 279, 0.061,-0.007&  91, 0.078,-0.052&  41, 0.050, 0.018&   4, 0.084, 0.000\\

\multicolumn{9}{c}{ $N_{\rm _{UV}}ugriz$, non-cata.}\\
 B&
1091, 0.107,-0.367&2337, 0.101,-0.361&1779, 0.100,-0.400& 765, 0.097,-0.434&
 236, 0.088,-0.450&  42, 0.082,-0.440&  14, 0.123,-0.457&   1, $\ \ \ \ \infty$,-0.337\\

\multicolumn{9}{c}{ $F_{\rm _{UV}}ugriz$, non-cata.}\\
 B&
1090, 0.063,-0.033&2334, 0.046, 0.013&1781, 0.051, 0.020& 776, 0.049, 0.010&
 280, 0.058,-0.011&  88, 0.083,-0.051&  42, 0.076, 0.002&   5, 0.066,-0.022\\

\hline
\end{tabular}
\label{Tab:perfor-z}
\end{table*}

\begin{table*}
\tiny \caption{ Similar to Table~\ref{Tab:perfor-z},
but for the bin of $g-r$ colour.}
\begin{tabular}{lcccccccc}
\hline
   & $g$$-$$r$=0.0$-$0.2 & $g$$-$$r$=0.2$-$0.4 & $g$$-$$r$=0.4$-$0.6 & $g$$-$$r$=0.6$-$0.8 & $g$$-$$r$=0.8$-$1.0 & $g$$-$$r$=1.0$-$1.2 & $g$$-$$r$=1.2$-$1.4 & $g$$-$$r$=1.4$-$1.6 \\
        & $N,\sigma_z,\langle\Delta z\rangle$ & $N,\sigma_z,\langle\Delta z\rangle$ & $N,\sigma_z,\langle\Delta z\rangle$ & $N,\sigma_z,\langle\Delta z\rangle$ & $N,\sigma_z,\langle\Delta z\rangle$ & $N,\sigma_z,\langle\Delta z\rangle$ & $N,\sigma_z,\langle\Delta z\rangle$ & $N,\sigma_z,\langle\Delta z\rangle$\\
\hline
 All&  17& 344&1109&1628&2050& 897& 230&  91( 160)\\

\multicolumn{9}{c}{ $ugriz$, non-cata.}\\
 A&
   1, $\ \ \ \infty$, -0.003&  64, 0.059,-0.041& 486, 0.067,-0.028&1212, 0.044,-0.011&
2018, 0.040, 0.015& 896, 0.046, 0.018& 228, 0.044, 0.008&  86, 0.081,-0.023\\
 B&
   3, 0.004, 0.007&  74, 0.041, 0.001& 680, 0.053,-0.010&1351, 0.045, 0.006&
2027, 0.040, 0.024& 896, 0.048, 0.014& 228, 0.052,-0.006&  87, 0.081,-0.051\\
 C&
   8, 0.030,-0.022&  94, 0.072,-0.048& 374, 0.075,-0.031& 836, 0.055,-0.037&
2025, 0.057,-0.030& 895, 0.062,-0.016& 228, 0.064,-0.020&  86, 0.083,-0.047\\
 C2&
  10, 0.049,-0.036&  86, 0.078,-0.051& 367, 0.076,-0.030& 835, 0.055,-0.036&
2023, 0.057,-0.029& 894, 0.063,-0.013& 228, 0.064,-0.018&  83, 0.079,-0.060\\
 D&
   1,$\ \ \ \ \infty$, -0.055&  38, 0.069,-0.066& 398, 0.073,-0.037&1160, 0.046,-0.020&
2023, 0.042, 0.012& 896, 0.049, 0.018& 228, 0.053, 0.010&  84, 0.082,-0.019\\
 D2&
   2, 0.003,-0.051&  51, 0.069,-0.040& 429, 0.071,-0.029&1195, 0.044,-0.013&
2018, 0.041, 0.013& 891, 0.049, 0.017& 197, 0.047, 0.008&  70, 0.090,-0.050\\

\multicolumn{9}{c}{ $F_{\rm _{UV}}N_{\rm _{UV}}ugriz$, non-cata.}\\
 A&
  16, 0.050,-0.045& 344, 0.053,-0.023&1104, 0.060,-0.013&1622, 0.054, 0.007&
2041, 0.048, 0.019& 897, 0.048, 0.020& 228, 0.052, 0.000&  90, 0.077,-0.036\\
 B&
  16, 0.059,-0.044& 344, 0.053,-0.005&1105, 0.062, 0.002&1621, 0.056, 0.014&
2040, 0.052, 0.021& 897, 0.052, 0.015& 228, 0.056,-0.010&  91, 0.081,-0.045\\
 C&
  17, 0.048,-0.037& 344, 0.066,-0.040&1105, 0.067,-0.033&1621, 0.069,-0.031&
2040, 0.063,-0.023& 897, 0.062,-0.025& 228, 0.074,-0.037&  89, 0.080,-0.068\\
 C2&
  17, 0.054,-0.043& 344, 0.068,-0.040&1105, 0.069,-0.033&1620, 0.069,-0.031&
2035, 0.064,-0.022& 897, 0.064,-0.021& 228, 0.074,-0.034&  89, 0.080,-0.071\\
 D&
  16, 0.053,-0.091& 344, 0.056,-0.046&1105, 0.061,-0.024&1621, 0.055,-0.003&
2039, 0.050, 0.014& 897, 0.047, 0.020& 228, 0.060, 0.004&  90, 0.092,-0.031\\
 D2&
  16, 0.054,-0.087& 344, 0.056,-0.044&1105, 0.062,-0.021&1621, 0.055, 0.001&
2037, 0.049, 0.014& 894, 0.047, 0.018& 224, 0.061, 0.001&  89, 0.087,-0.047\\

\multicolumn{9}{c}{ $N_{\rm _{UV}}ugriz$, non-cata.}\\
 B&
  17, 0.084,-0.392& 342, 0.086,-0.359&1105, 0.100,-0.321&1619, 0.109,-0.350&
1980, 0.079,-0.410& 880, 0.080,-0.451& 207, 0.071,-0.506&  61, 0.112,-0.485\\

\multicolumn{9}{c}{ $F_{\rm _{UV}}ugriz$, non-cata.}\\
 B&
  17, 0.102,-0.070& 339, 0.056,-0.014&1094, 0.065,-0.011&1563, 0.056, 0.004&
2032, 0.045, 0.020& 896, 0.049, 0.013& 228, 0.048,-0.006&  87, 0.073,-0.046\\

\hline
\end{tabular}
\label{Tab:Tab:perfor-gr}
\end{table*}

\begin{table*}
\tiny \caption{ Similar to Table~\ref{Tab:perfor-z},
but for the $r$-magnitude bin.}
\begin{tabular}{lcccccccc}
\hline
   & $r$=15.5$-$16.0 & $r$=16.0$-$16.5 & $r$=16.5$-$17.0 & $r$=17.0$-$17.5 & $r$=17.5$-$18.0 & $r$=18.0$-$18.5 & $r$=18.5$-$19.0 & $r$=19.0$-$19.5 \\
        & $N,\sigma_z,\langle\Delta z\rangle$ & $N,\sigma_z,\langle\Delta z\rangle$ & $N,\sigma_z,\langle\Delta z\rangle$ & $N,\sigma_z,\langle\Delta z\rangle$ & $N,\sigma_z,\langle\Delta z\rangle$ & $N,\sigma_z,\langle\Delta z\rangle$ & $N,\sigma_z,\langle\Delta z\rangle$ & $N,\sigma_z,\langle\Delta z\rangle$\\
\hline
All &
(130) 149& 275& 478& 929&1498&1952& 834&  96 (187) \\

\multicolumn{9}{c}{ $ugriz$, non-cata.}\\
 A&
 117, 0.041,-0.005& 212, 0.051,-0.008& 358, 0.043, 0.003& 660, 0.047,-0.002&
1062, 0.050, 0.002&1536, 0.050, 0.008& 825, 0.046, 0.010&  94, 0.079,-0.011\\
 B&
 124, 0.041, 0.001& 225, 0.045, 0.005& 383, 0.036, 0.014& 726, 0.046, 0.009&
1171, 0.049, 0.012&1674, 0.046, 0.016& 825, 0.049, 0.011&  94, 0.077,-0.028\\
 C&
 122, 0.061,-0.039& 202, 0.061,-0.027& 328, 0.057,-0.019& 583, 0.060,-0.028&
945, 0.062,-0.029&1345, 0.060,-0.027& 809, 0.059,-0.031&  94, 0.077,-0.037\\
 C2&
 120, 0.059,-0.039& 201, 0.062,-0.029& 326, 0.058,-0.018& 579, 0.061,-0.025&
 944, 0.062,-0.028&1340, 0.061,-0.026& 809, 0.060,-0.030&  92, 0.076,-0.038\\
 D&
 115, 0.048,-0.010& 208, 0.054,-0.008& 351, 0.048,-0.001& 627, 0.051,-0.003&
1013, 0.054,-0.002&1478, 0.053, 0.003& 819, 0.049, 0.008&  94, 0.085,-0.007\\
 D2&
 118, 0.047,-0.010& 213, 0.051,-0.013& 355, 0.043, 0.001& 638, 0.049,-0.001&
1035, 0.051, 0.002&1491, 0.049, 0.007& 803, 0.049, 0.008&  78, 0.083,-0.022\\

\multicolumn{9}{c}{ $F_{\rm _{UV}}N_{\rm _{UV}}ugriz$, non-cata.}\\
 A&
 149, 0.052,-0.015& 274, 0.054,-0.009& 478, 0.056, 0.001& 927, 0.049, 0.004&
1489, 0.054, 0.006&1944, 0.053, 0.015& 831, 0.049, 0.016&  96, 0.067,-0.018\\
 B&
 149, 0.054,-0.015& 273, 0.053,-0.005& 478, 0.056, 0.004& 927, 0.049, 0.011&
1489, 0.054, 0.014&1945, 0.057, 0.021& 831, 0.053, 0.014&  96, 0.075,-0.025\\
 C&
 149, 0.069,-0.040& 273, 0.070,-0.029& 478, 0.065,-0.021& 927, 0.065,-0.027&
1490, 0.065,-0.028&1944, 0.064,-0.024& 831, 0.062,-0.037&  96, 0.086,-0.060\\
 C2&
 149, 0.074,-0.039& 270, 0.072,-0.032& 476, 0.067,-0.020& 927, 0.066,-0.027&
1490, 0.066,-0.027&1944, 0.065,-0.023& 831, 0.063,-0.035&  96, 0.086,-0.061\\
 D&
 149, 0.060,-0.023& 272, 0.059,-0.017& 477, 0.060,-0.007& 928, 0.053,-0.004&
1490, 0.057,-0.002&1944, 0.055, 0.007& 831, 0.048, 0.013&  96, 0.081,-0.018\\
 D2&
 149, 0.057,-0.023& 271, 0.056,-0.014& 476, 0.059,-0.006& 928, 0.054,-0.003&
1488, 0.056, 0.000&1944, 0.054, 0.009& 827, 0.050, 0.013&  95, 0.080,-0.028\\

\multicolumn{9}{c}{ $N_{\rm _{UV}}ugriz$, non-cata.}\\
 B&
 137, 0.097,-0.381& 258, 0.099,-0.374& 456, 0.090,-0.367& 923, 0.096,-0.366&
1482, 0.102,-0.375&1927, 0.106,-0.378& 814, 0.095,-0.441&  73, 0.083,-0.516\\

\multicolumn{9}{c}{ $F_{\rm _{UV}}ugriz$, non-cata.}\\
 B&
 143, 0.052,-0.006& 268, 0.052, 0.000& 471, 0.057, 0.005& 918, 0.054, 0.004&
1472, 0.055, 0.007&1913, 0.055, 0.010& 830, 0.051, 0.011&  94, 0.070,-0.026\\

\hline
\end{tabular}
\label{Tab:perfor-r}
\end{table*}

\section{Dependence of morphology accuracy on redshift, colour and magnitude.}
{ In Figs.C1-C3 we illustrate the evolutions of $N_i$, $|\langle \Delta T_i \rangle|$ and $\sigma _{T_i}$ as a function of spectroscopic redshift $z_{\rm spec}$, $g-r$ colour and $r$-magnitude in removing catastrophic identifications for Models A-D2. This part is based on morphological galaxy sample.}

\begin{figure*}
\centering{
\includegraphics[bb= 47 443 534 743,height= 8.5cm,width=18.5cm,clip,angle=0]{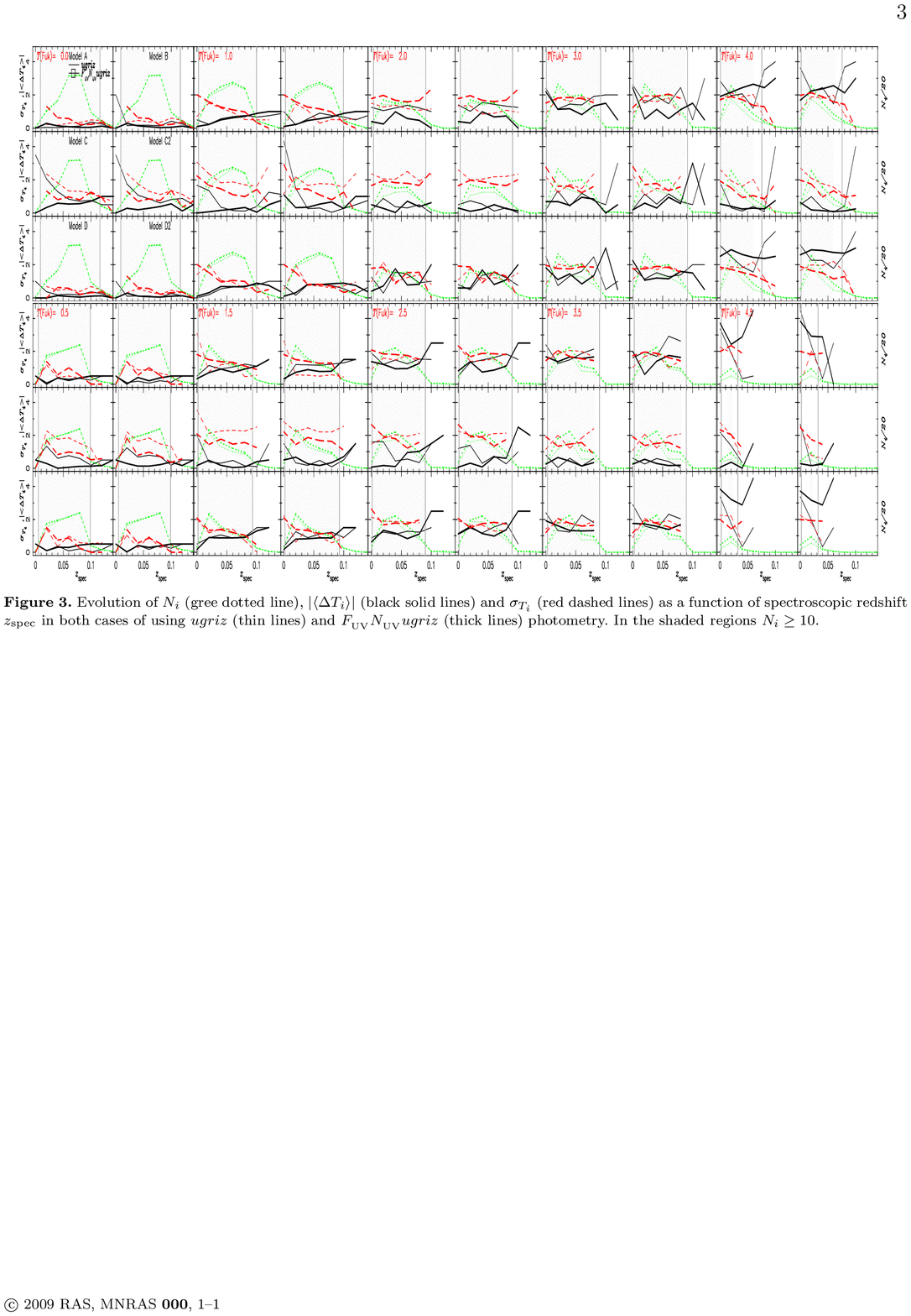}
}
\caption{ Evolutions of $N_i$ (green dotted line),
$|\langle \Delta T_i \rangle|$ (black solid lines) and $\sigma
_{T_i}$ (red dashed lines) as a function of spectroscopic redshift
$z_{\rm spec}$ in both cases of using $ugriz$ (thin lines) and
$F_{_{\rm UV }}N_{_{\rm UV }}ugriz$ (thick lines) photometry.
In the shaded regions $N_i \ge 10$.
}
\label{aFig:z}
\end{figure*}
%

\begin{figure*}
\centering{
\includegraphics[bb= 47 445 534 743,height= 8.5cm,width=18.5cm,clip,angle=0]{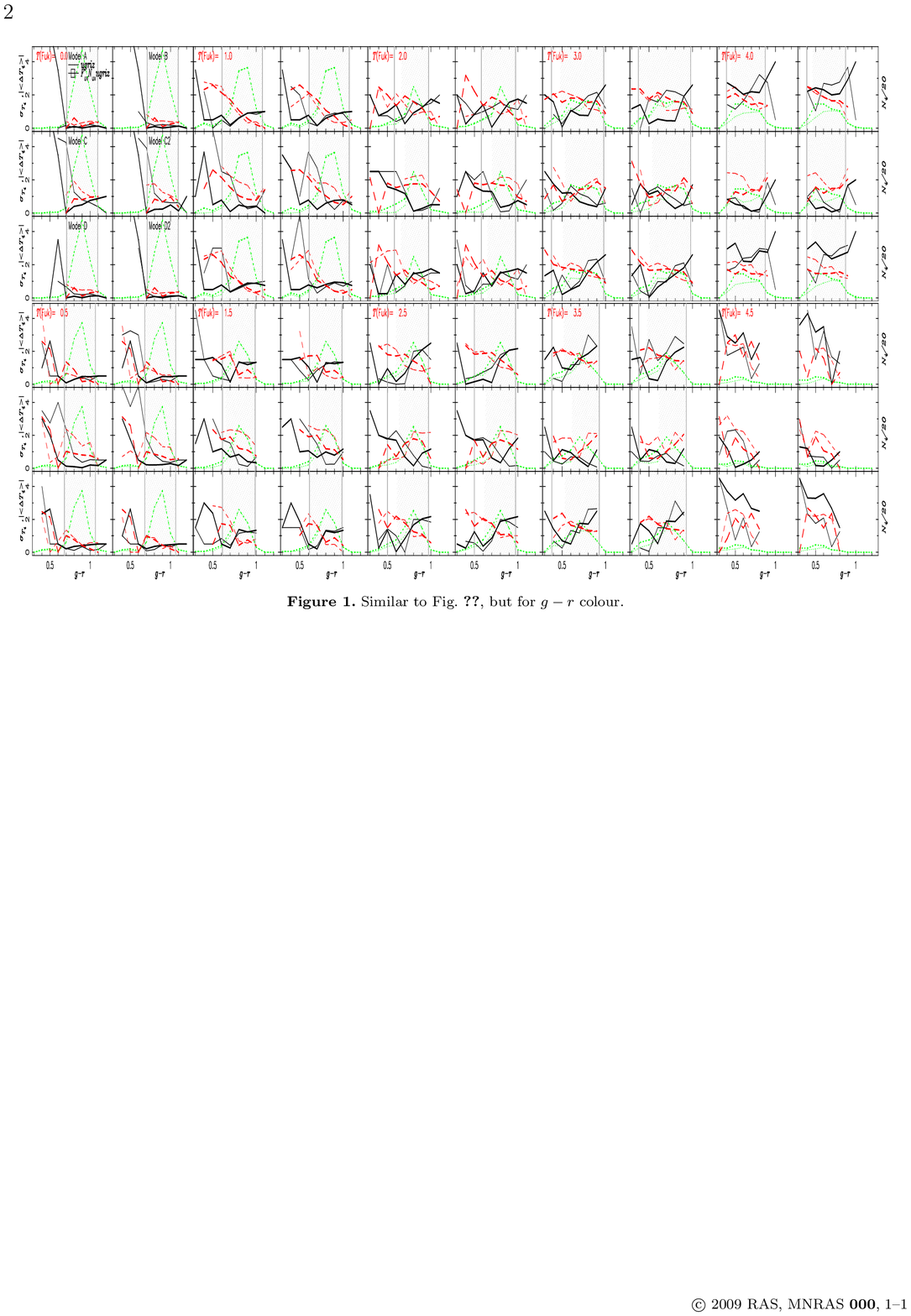}
}
\caption{Similar to Fig.~\ref{aFig:z}, but as a function of $g-r$ colour.}
\label{aFig:gr}
\end{figure*}
%

\begin{figure*}
\centering{
\includegraphics[bb= 47 445 534 743,height= 8.5cm,width=18.5cm,clip,angle=0]{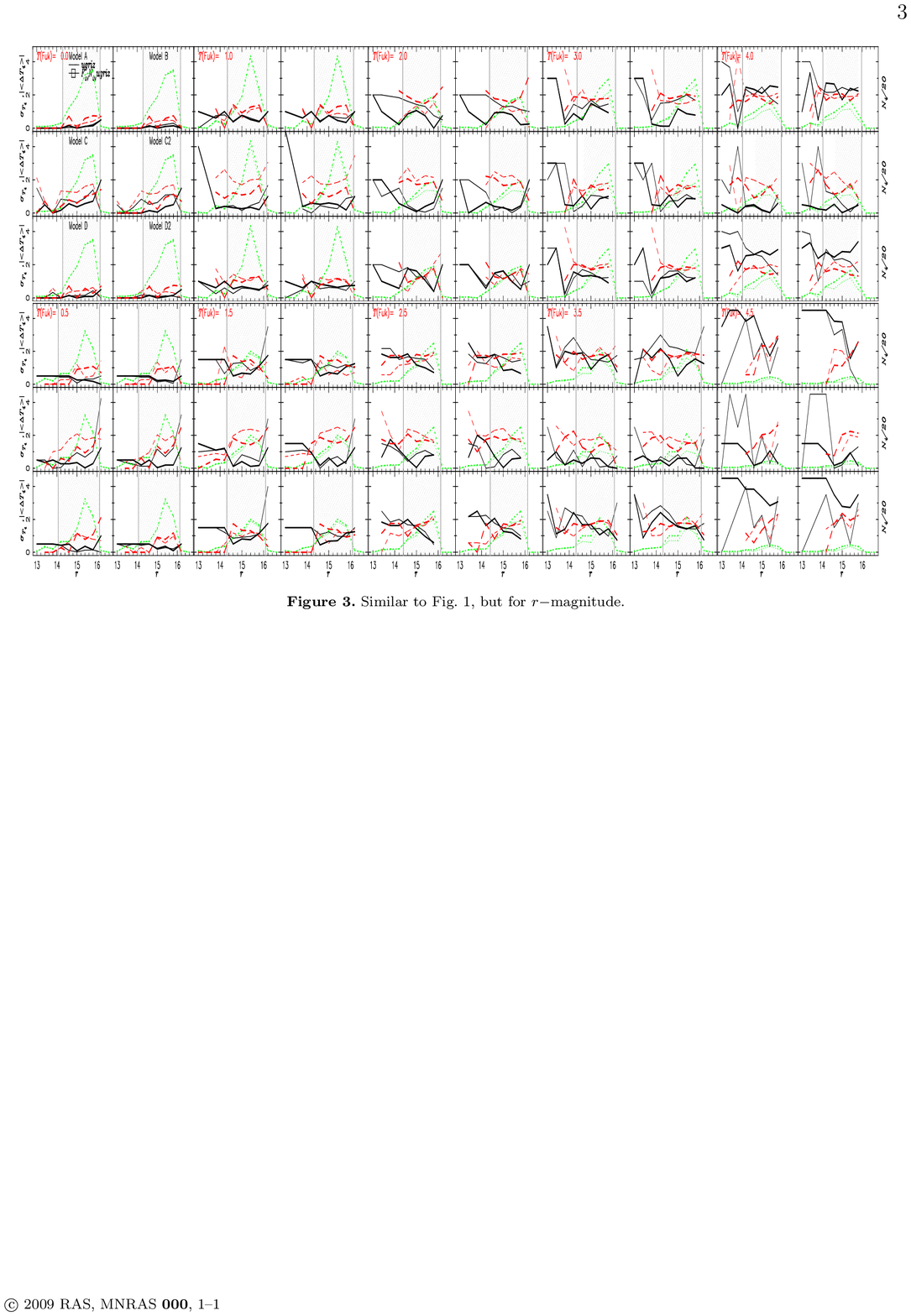}
}
\caption{ Similar to Fig.~\ref{aFig:z}, but as a function of $r-$magnitude.}
\label{aFig:r}
\end{figure*}

\bsp
\label{lastpage}
\end{document}